\documentstyle[onecolumn,epsfig]{mn}
\newcommand{\mincir}{\raise
-2.truept\hbox{\rlap{\hbox{$\sim$}}\raise5.truept
\hbox{$<$}\ }}
\newcommand{\magcir}{\raise
-2.truept\hbox{\rlap{\hbox{$\sim$}}\raise5.truept
\hbox{$>$}\ }}
\newcommand{\minmag}{\raise-2.truept\hbox{\rlap{\hbox{$<$}}\raise
6.truept\hbox{$>$}\ }}
\newcommand{\be}{\begin{equation}}
\newcommand{\ee}{\end{equation}}
\newcommand{\ba}{\begin{eqnarray}}
\newcommand{\ea}{\end{eqnarray}}
\newcommand{\brr}{\begin{array}}

\newcommand{\err}{\end{array}}
\newcommand{\bc}{\begin{center}}
\newcommand{\ec}{\end{center}}

\title{The growth of structure in the intergalactic medium}
\author[S. Matarrese and R. Mohayaee]
{Sabino Matarrese $^1$
and Roya Mohayaee $^{1,2}$
\\
$1$ Dipartimento di Fisica `Galileo Galilei', Universit\`a di
Padova,
via Marzolo 8, I-35131 Padova, Italy \\
$^2$ Dipartimento di Fisica, Universit\`a di Roma
`La Sapienza', P.le Aldo Moro 5, 00185, Roma, Italy\\
email: matarrese@pd.infn.it\\
email: roya@pd.infn.it
}
\begin{document}

\maketitle

\begin{abstract}
A {\it stochastic adhesion} model is introduced, with the
purpose of describing the formation and evolution of mildly
nonlinear structures, such as sheets and filaments, in the
intergalactic medium (IGM), after hydrogen reionization. The model
is based on replacing the overall force acting on the baryon fluid
-- as it results from the composition of local gravity, pressure
gradients and Hubble drag -- by a mock external force,
self-consistently calculated from first-order perturbation theory.
A small kinematic viscosity term prevents shell-crossing on small
scales (which arises because of the approximate treatment of
pressure gradients). The emerging scheme is an extension of the
well-known adhesion approximation for the dark matter dynamics,
from which it only differs by the presence of a small-scale
`random' force, characterizing the IGM. Our algorithm is the ideal
tool to obtain the skeleton of the IGM distribution, which is
responsible for the structure observed in the low-column density
Ly$\alpha$ forest in the absorption spectra of distant quasars.

\end{abstract}


\section{Introduction}

The analysis of the spectra of distant QSOs blueward of the
Ly$\alpha$ emission, the so-called Ly$\alpha$ forest, has revealed
the presence of large coherent structures in the cosmic
distribution of the neutral hydrogen, left over by the
reionization process (e.g. Rauch 1998 and references therein). The
low-column density ($N_{\rm HI} \leq 10^{14.5}$ cm$^{-2}$) Ly$\alpha$
forest is thought as being generated
by unshocked gas in voids and mildly overdense fluctuations of the
photoionized intergalactic medium (IGM). The IGM is, in turn,
expected to trace the underlying dark matter (DM) distribution on
large scales, while the thermal pressure in the gas smears
small-scale DM nonlinearities. According to this picture, the
large coherent spatial structures in the IGM, which are probed by
quasar absorption spectra, reflect the underlying network of
filaments and sheets in the large-scale DM distribution (e.g.
Efstathiou, Schaye \& Theuns 2000), the so-called `cosmic web'
(Bond, Kofman \& Pogosyan 1996). Thanks to this simple picture, a
number of semi-analytical, or pseudo-hydrodynamical
particle-mesh codes have been developed, which all aim at
simulating the IGM distribution, with enough resolution to predict
the main features of the Ly$\alpha$ forest (e.g. Bi \& Davidsen 1997; Croft
et al. 1998; Gnedin \& Hui 1996, 1998; Hui, Gnedin \& Zhang 1997; 
Petitjean, M\"ucket \& Kates 1995; Gazta\~naga \&
Croft 1999; Meiksin \& White 2000; Viel et al. 2001). This
physical interpretation of the Ly$\alpha$ forest is, moreover,
largely corroborated by numerical simulations based on fully
hydrodynamical codes (e.g. Cen et al. 1994; Zhang, Anninos \&
Norman 1995, 1997; Miralda-Escud\'e et al. 1996; Hernquist et al.
1996; Theuns et al. 1998).

Although the above scenario captures the main physical processes
which determine the coarse-grained dynamics of the baryons, it
neglects a number of fine-grained details which distinguish the
baryon distribution from the underlying DM one. There has been,
recently, growing observational and theoretical interest on the
low-column density Ly$\alpha$ forest, as an important indicator
of the state of the Universe in the redshift range $2 - 4$. We
then believe it has become necessary to develop new and more
sophisticated techniques, enabling to understand the IGM dynamics
and accurately simulate its clustering, without resorting to the
full machinery of hydro-simulations. The model presented here aims
at providing an accurate scheme for the growth and evolution of
mildly nonlinear structures in the baryon gas, accounting for the
detailed thermal history of the IGM.

The {\it stochastic adhesion} model introduced here is based on
applying the {\it forced Burgers equation} of nonlinear diffusion
(Forster, Nelson \& Stephen 1977; Kardar, Parisi \& Zhang 1986; 
Barab\'asi \& Stanley 1995; E et al. 1997; Frisch \& Bec 2000 
and references therein) to the
cosmological framework, where it can be given the general form
\be
{D{\bf u}({\bf x},\tau) \over D \tau} = \nu \nabla^2 {\bf u}({\bf
x},\tau) - \nabla \eta({\bf x},\tau) \;.
\label{genforcedburgers}
\ee
Here $\bf x$ are comoving coordinates, the time variable $\tau$ is
chosen to coincide with the cosmic scale-factor $a$ (although, for
the pure DM evolution, a better choice would be the linear growth
factor of density fluctuations), $\bf u \equiv  d{\bf x} /d \tau$ is the
(suitably rescaled) peculiar velocity field, which is here assumed
to be irrotational, ${\bf u} = \nabla \Phi$, and $D / D\tau \equiv
\partial/\partial \tau + {\bf u} \cdot \nabla$ denotes the convective
time-derivative. The coefficient of kinematical viscosity, $\nu$,
is always assumed to be small. In standard applications of the
forced Burgers equation, $\eta$ is a random external potential.
For a general cosmological fluid, such as the dark matter or the
baryons, $\eta = (3 /2 a) (\Phi + \varphi + w)$, where $\varphi$
is the local gravitational potential (up to an appropriate
rescaling, whose precise form will be given in Section 2), which
must be consistently determined via the Local Poisson equation,
and $w$ is the specific enthalpy of the fluid (up to the same
rescaling), which vanishes in the DM case. Therefore, our
potential $\eta$ in the RHS of eq. (\ref{genforcedburgers}) is far
from being `external', as it is, for instance, dynamically related
to the velocity field which appears on the LHS of the same
equation.

The main idea of this paper is that, if $\eta$ is given an
approximate expression, e.g. by using the results of perturbation
theory, it can be legitimately treated as a truly external random
potential. The choice of variables in eq. (\ref{genforcedburgers})
is such that, if this approximation technique is applied to the DM
evolution, to first order it reduces to the well-known adhesion
model (Gurbatov, Saichev \& Shandarin, 1985, 1989; Kofman \&
Shandarin 1988), which was introduced in cosmology to extend the
validity of the Zel'dovich approximation (Zel'dovich 1970) beyond
the epoch of first caustic formation. In the DM case, therefore,
only a second-order calculation would produce a non-vanishing
external force, whose presence would then affect rather small
scales. Quite different is the case of the collisional baryon
component, where, already at first order in perturbation theory
(both Eulerian and Lagrangian) a non-zero contribution to $\eta$
generally comes out: it is originated from the unbalanced
composition of Hubble drag, local gravity and gas pressure, thus
representing a genuine baryonic feature.

The presence of the kinematical viscosity term requires some
explanation. Its role in the present context is twofold. First, it
prevents the formation of multi-streams, which are well-known to
affect the DM dynamics, but may also appear as a spurious effect
in the collisional case, when pressure gradients are given an
approximate form in terms of linear theory. Second, it allows to
transform the problem into a linear one, through the so-called
Hopf-Cole substitution (e.g. Burgers 1974) ${\bf u} = -2 \nu
\nabla_{\bf x} \ln {\cal U}$, where the `expotential' ${\cal U}$
obeys the {\it random heat}, linear diffusion, equation
\be
{\partial {\cal U}({\bf x},\tau) \over \partial \tau} = \nu
\nabla^2 {\cal U}({\bf x},\tau) + {\eta({\bf x},\tau) \over 2\nu}
{\cal U}({\bf x},\tau) \;, \label{genrandomheat}
\ee
whose solution is expressible in terms of path-integrals
(e.g. Feynman \& Hibbs 1965). As we will see in Sections 5 and 6,
a suitable approximation technique, valid in the small $\nu$ case,
allows to give the velocity field a simple and finite form, more
convenient for practical applications.

The forced Burgers equation has been used to describe a variety of
different physical problems, ranging from interfacial growth
in condensed matter physics, where it is known as the KPZ model
(Kardar, Parisi \& Zhang 1986), to fully developed turbulence
(e.g. Bouchaud, M\'ezard \& Parisi 1995). Later in the paper we
will come back to this interesting connection and discuss both the
analogies and the peculiarities of the cosmological application of
this equation.

The stochastic adhesion approximation provides an analytical
description of the generation, and subsequent merging of shocks,
which give rise to the thin network of filaments and sheets in
the IGM spatial distribution. Their existence is clearly observed
in the spectra of high-redshift QSOs, e.g. through the presence
of common absorption lines in the Ly$\alpha$ forest of multiple QSOs, 
with lines of sight separated by several comoving Mpc at redshifts 
$z \approx 2 - 4$ (e.g. Rauch 1998 and references therein). An
important property of our model is that it allows to draw the
skeleton of the IGM distribution through a straightforward
extension of the geometrical technique applied in the free
adhesion model (e.g. Sahni \& Coles 1995 and references therein).
The present paper will be mostly devoted to provide the physical
and mathematical bases for our stochastic adhesion model. 
Simulations of the IGM large-scale structure will be obtained 
only from the simplified inviscid ($\nu=0$) model.   
In a subsequent paper we will implement our algorithm to
produce numerical simulations of the IGM distribution
and to study the statistical properties of the IGM density and
velocity fields.  Let us stress that approximation schemes like 
the present one can be particularly useful, as they allow to 
better account for the cosmic variance of large-scale modes, which is 
poorly probed (especially at low redshifts) by the
existing hydro-simulations, which are forced to adopt small 
computational boxes to increase the small-scale resolution [e.g.  
the discussion in (Viel et al. 2001)]. Even more interesting is the 
possibility to combine our scheme with a hydro-code, using the former 
to provide the large-scale skeleton of the IGM and the latter 
to achieve the required resolution on small scales. 
 
The approach most closely related to ours is that recently
proposed by Jones (1996, 1999), which aims at modelling the
nonlinear clustering of the baryonic material. In Jones' model,
however, a different set of variables is adopted, which does not
allow a direct comparison neither with the present scheme, nor
with the Zel'dovich and the adhesion approximations, in the
collisionless limit. The most important difference is that, in
Jones'  model, the 
external random term is identified with the local gravitational
potential $\varphi$, which is treated as an `external' one, as it
is essentially generated by the dominant DM component; moreover,
no explicit account for the gas pressure is given. In our model
instead, the external potential $\eta$ is obtained by linearly
approximating the composition of Hubble drag, local gravity and
thermal pressure which act on the IGM fluid elements; in our case
$\eta$ is an external potential because it is determined by a
convolution of the initial gravitational potential with the IGM
linear filter.

It should be clear from this introduction that the forced Burgers
equation might have wider applications in the cosmological
structure formation problem. Its relevance (in terms of the
closely related random heat equation) in the cosmological
framework has been first advocated by Zel'dovich and collaborators
in the mid eighties (Zel'dovich et al. 1985, 1987), as a means to
describe the possible origin of {\it intermittency} in the matter
distribution. The intermittency phenomenon consists in the
appearance of rare high peaks, where most of the matter is
concentrated, separated by vast regions of reduced intensity. From
the statistical point of view, intermittency in a stochastic
process is signalled by an anomalous scaling of e.g. structure
functions (moments of velocity increments):
moments of order $p>2$, made dimensionless by suitable powers of
the second-order moment, grow without bound on small
scales (e.g. G\"artner J., Molchanov 1990; Frisch 1995). 
This may be viewed as increased
non-Gaussianity on small scales, which, in Fourier space,
appears as a slow decrease in the amplitude of Fourier modes
with increasing wavenumber and by a definite phase relation
between them (Zel'dovich et al. 1985, 1987). Actually,
the occurrence of this form of intermittency,
which seems too extreme and far from our present understanding of
the large-scale structure of the Universe, is usually obtained
under special properties of the noise and only appears at
asymptotically late times. Much more interesting for the structure
formation problem is a second phenomenon, called {\it intermediate
intermittency}, also described by the forced Burgers equation,
which consists in the formation of a cellular, or network
structure, with ``thin channels of raised intensity (the rich
phase), separating isolated islands of the poor phase" (Zel'dovich
et al. 1985). This second phenomenon is expected to arise as an
intermediate asymptotic situation.

Can one take advantage of the description of the structure
formation process in terms of intermediate intermittency to
predict the specific non-Gaussian statistics which characterizes
the nonlinear density field? This is, we believe, a challenging
issue, which would deserve further analysis.

The prototype distribution that describes intermittency is the
Lognormal one, which naturally arises in multiplicative processes,
through the action of the Central Limit Theorem (e.g. Shimizu \&
Crow 1988).
In the DM case, Coles \& Jones (1991) proposed
that a local Lognormal mapping of the linear density field can
describe the nonlinear evolution of structures in the Universe.
Detailed comparison with N-body simulations showed that this model
fits very well the bulk of the probability density function (PDF)
of the mass density field in N-body simulations, for moderate
values of the {\it rms} overdensity $\sigma$ (e.g. Bernardeau \&
Kofman 1995). Where the Lognormal fails is in reproducing the
correct PDF for the strong clustering regime ($\sigma\gg 1$), as
well as the high- and low-density tails of the PDF, even on mildly
nonlinear scales. Moreover, the predicted spatial pattern is too
clumpy and poorly populated of extended structures to reproduce
that of simulations (Coles, Melott \& Shandarin 1993). A `skewed'
Lognormal PDF is proposed by Colombi (1994), to follow the
transition from the weakly to the highly nonlinear regime.

Quite noticeably, the Lognormal model has also been used in
connection with the IGM dynamics. Indeed, the semi-analytical
model proposed by Bi and collaborators (Bi 1993; Bi et al. 1995;
Bi \& Davidsen 1997; see also Feng \& Fang 2000; Roy Choudhury,
Padmanabhan \& Srianand 2000; Roy Choudhury, Srianand \&
Padmanabhan 2000; Viel et al. 2001), to simulate the low-column
density Ly$\alpha$ forest, is based on a local Lognormal model,
similar to the one of Coles \& Jones (1991). Comparison with the
results of more refined techniques has shown that it provides a
good fit of the column density distribution in a wide range of
values, but it tends to underestimate the abundance of lower
column density systems (Hui, Gnedin \& Zhang 1997). Moreover, the
Lognormal model for the IGM tends to produce an excess of
saturated absorption lines in the simulated transmitted flux,
compared with real QSO spectra.
\\

The plan of the paper is as follows. In Section 2 we review the
equations which govern the Newtonian dynamics of a two-component
fluid of dark matter (DM) and baryons, in the expanding Universe;
these are solved in Section 3 at the Eulerian linear level and
under fairly general assumptions on the baryon equation of state.
This allows to obtain the IGM filter connecting the linear baryon
density fluctuations to the DM ones. A simplified version of our
model is presented in Section 4: it enables one to follow the
combined dark matter and baryon dynamics on weakly nonlinear
scales, within the laminar regime. In Section 4 we also give the
first-order Lagrangian solution for the baryon dynamics, which is
then used to perform numerical simulations of the IGM
distribution. In Section 5 we discuss the modifications introduced
in the baryon dynamics by our improved final model, where we add a
kinematic viscosity term, to avoid the occurrence of
shell-crossing singularities (arising also in the baryon fluid,
because of the approximate treatment of pressure gradients). This
leads to our stochastic adhesion model for the dynamics of the
intergalactic medium. According to this model, the approximate IGM
dynamics is governed by the forced Burgers equation for the baryon
peculiar velocity field, whose solution can be expressed as a
path-integral. In the physically relevant limit of small
viscosity, a solution can be found through the standard
saddle-point technique. In Section 6 we discuss how to implement
our solution in terms of the first-order Lagrangian particle
trajectories previously obtained. A geometrical algorithm is also
outlined, which allows to draw the skeleton of the IGM
distribution, given a realization of the gravitational potential
and the linear IGM filter. A preliminary analysis of the
statistical properties of the density field obtained through the
stochastic adhesion model is given in Section 7. The concluding
Section 8 contains a brief discussion on the possible applications
of our model.

\section{Dynamics of dark matter and baryons in the expanding Universe}

The Newtonian dynamics of a self-gravitating two-component fluid,
made of collisionless dark matter and collisional baryonic gas is
governed by the continuity, Euler and Poisson equations. The
continuity equation for the dark matter component (indicated by a
subscript ${\rm DM}$) reads
\be
{\partial\rho_{\rm DM}\over\partial t} +
3 H \rho_{\rm DM} + \nabla \cdot \left(\rho_{\rm DM} {\bf v}_{\rm
DM} \right)  = 0 \;,
\ee
where $\rho$ is the mass density, ${\bf v} \equiv a d {\bf x}/ dt$
the peculiar velocity and $H$ the Hubble parameter at time $t$;
the Euler equation reads
\be
{\partial (a {\bf v}_{\rm DM})
\over\partial t} + \left({\bf v}_{\rm DM} \cdot \nabla \right){\bf
v}_{\rm DM}  = - \nabla \phi \;,
\ee
where $\phi$ is the peculiar gravitational potential. For the
baryon fluid (indicated by a subscript $b$), we have
\be
{\partial\rho_b\over\partial t} + 3 H \rho_{\rm b} + \nabla \cdot
\left(\rho_{\rm b} {\bf v}_{\rm b}\right) = 0 \;,
\ee
and
\be
{\partial (a {\bf v}_{\rm b}) \over\partial t} + \left({\bf
v}_{\rm b} \cdot \nabla \right) {\bf v}_{\rm b} = - \nabla\phi -
\frac{\nabla p_{\rm b}}{\rho_{\rm b}}\;,
\ee
where $p$ is the pressure. The peculiar gravitational potential
obeys the Poisson equation $\nabla^2 \phi = 4 \pi G a^2 \delta
\rho$, where the mass-density fluctuation $\delta\rho$ takes
contribution both from dark matter and baryons. Let then $f_{\rm
DM}$ and $f_{\rm b} = 1 - f_{\rm DM}$ be the mean mass fraction of
these two types of matter. We have
\be
\nabla^2 \phi = \frac{3}{2}
H_0^2 \Omega_{0{\rm m}} \left(f_{\rm DM}\delta_{\rm DM} + f_b
\delta_{\rm b}\right) \;,
\label{twocomp}
\ee
where $H_0$ is the Hubble constant, $\Omega_{0{\rm m}}$ is the
closure density of non-relativistic matter (both dark matter and
baryons) today, $\delta_{\rm DM}$ and $\delta_{\rm b}$ are the
fractional DM and baryon overdensities. In the analytical
calculations which follow we will neglect, for simplicity, the
self-gravity of the baryons, i.e. we will put $f_{\rm DM}=1$.

To close the system we need the IGM equation of state, which can
be taken of the polytropic form
\be
p_b = \frac{\rho_{\rm b} k_B
T}{\mu m_p} = {\overline{\rho}_{\rm b} k_B T_0 (1+\delta_{\rm
b})^\gamma \over \mu m_p} \;,
\ee
where $k_B$ is Boltzman's constant, $\gamma$ the adiabatic index,
$\mu$ the mean molecular weight (for fully ionized gas with
primordial abundances it is about $0.6$), $m_p$ the proton mass
and $\overline{\rho}_{\rm b}$ the baryon mean density. In writing
the pressure term we have assumed the power-law
temperature-density relation $T=T_0(1+\delta_b)^{\gamma-1}$ where
$T_0 = T_0(z)$ is the IGM temperature at mean density, at redshift
$z$ (e.g. Hui \& Gnedin 1997; Schaye et al. 1999; McDonald et al. 2000).
This is adequate for low to moderate baryon overdensity
($\delta_{\rm b} < 10$), where the temperature is locally
determined by the interplay between photoheating by the UV
background and adiabatic cooling due to the Universe expansion.
The underlying assumption for this `equation of state' is a tight
local relation between the temperature and the baryon density,
which is only true for unshocked gas (e.g. Efstathiou et al.
2000).

In what follows it will prove convenient to change time variable
from the cosmic time $t$ to the scale-factor $a$ (e.g. Shandarin
\& Zel'dovich 1989; Matarrese et al. 1992; Sahni \& Coles 1995).
This defines new peculiar velocity fields ${\bf u} \equiv d{\bf
x}/d a = {\bf v}/(a^2 H)$. Let us also introduce the dimensionless
comoving densities $\rho  /\overline \rho = 1 + \delta$ and a
scaled gravitational potential $\varphi = 2 \phi/(3a^3 H^2)$.
For the redshift range of interest here ($z \approx 2 - 4$), one can write
$a^3 H^2 \approx H_0^2 \Omega_{0{\rm m}}$. 
A more exact treatment of the DM component would
require the use of the growing mode of linear density
perturbations, $D_+(t)$, as time variable (e.g. Gurbatov et al.
1989; Catelan et al. 1995); once again, for the range of redshifts
of interest here one can safely assume $D_+(t) \propto a(t)$. In
the Einstein-de Sitter case the present treatment becomes exact.

We have the following set of equations for the DM component
\be
{D {\bf u}_{\rm DM} \over D a}  =  - {3 \over 2 a} \left({\bf
u}_{\rm DM} +
\nabla \varphi \right) \qquad ;\qquad
{D \delta_{\rm DM} \over D a}  =  - (1 + \delta_{\rm DM}) \nabla \cdot
{\bf u}_{\rm DM} \;,
\ee
where  $D / Da \equiv \partial / \partial a + {\bf u} \cdot \nabla$
denotes the convective, or Lagrangian, derivative w.r.t. our new time
variable $a(t)$. For the baryons, we have
\ba
{D {\bf u}_b \over D a} & = & - {3 \over 2 a} \left[ {\bf u}_b + \nabla
\varphi + \left( {2 \gamma k_B T_0 \over 3 H_0^2 \Omega_{0{\rm m}} \mu m_p}
\right) (1 + \delta_{\rm b})^{\gamma -2} \nabla \delta_{\rm b}\right] \\
{D \delta_{\rm b} \over D a} & = & - (1 + \delta_{\rm b}) \nabla
\cdot {\bf u}_{\rm b} \;.
\ea
The Poisson equation also gets simplified:
\be
\nabla^2 \varphi =
{\delta_{\rm DM} \over a} \;.
\ee

\section{Linear theory}

\subsection{Dark matter in the linear regime}

Let us start by writing the above set of equations for the DM
component in the (Eulerian) linear approximation. The continuity
and Euler equations, respectively, simplify to
\be
{\dot \delta}_{\rm DM}  =  -
\nabla \cdot {\bf u}_{\rm DM}
\qquad;\qquad
{\dot {\bf u}}_{\rm DM}  = 
-{3\over 2 a} \left( {\bf u}_{\rm DM}+ \nabla\varphi \right) \;,
\ee
where, from now on, dots will denote partial differentiation w.r.t.
the scale-factor $a$.
The solutions are well known, and we will simply report them here. Keeping
only the growing mode terms we have
\be
\delta_{\rm DM}({\bf x},a)= a \delta_0({\bf x})
\qquad;\qquad
\varphi({\bf x},a)=\varphi_0({\bf x})
\qquad;\qquad
{\bf u}_{\rm DM}({\bf x},a)= - \nabla \varphi_0({\bf x}) \;,
\ee
where $\delta_0=\nabla^2\varphi_0$.

\subsection{Baryons in the linear regime}

In order to make a similar analysis for the baryons we need to specify the
time (or redshift) dependence of the baryon mean temperature $T_0$, which
generally depends upon the thermal history of the IGM, as well as
on the spectral shape of the UV background.\footnote{Similar reasonings would
actually also apply to the adiabatic index $\gamma$, which we will, however,
approximate as being constant in what follows.}
We will here consider a simple, but fairly general law of the
type $T_0(z) \propto (1+z)^\alpha$ which will allow to get exact solutions
of the linearized baryon equations.

At high redshifts, before decoupling, the mean baryon temperature
drops like $T_0\propto (1+z)$; when Compton scattering becomes
inefficient adiabatic cooling of the baryons implies $T_0 \propto
(1+z)^2$, which makes it practically vanish before reionization
[this will justify our initial conditions for the evolution of the
baryon overdensity in eqs. (\ref{initial})]. As the Universe
reionizes, the IGM temperature rises and a different redshift
dependence takes place. Various types of dependence have been
considered in the literature. The linear law $T_0 \propto (1+z)$
is often assumed for simplicity. According to Miralda-Escud\'e \&
Rees (1994) and Hui \& Gnedin (1997), long after hydrogen
reionization has occurred the diffuse IGM settles into an
asymptotic state where adiabatic cooling is balanced by
photoheating, leading to the power-law $T_0 \propto (1+z)^\alpha$,
with $\alpha=1/1.76$. Much steeper an exponent, $\alpha=1.7$, has
been recently adopted by Bryan \& Machacek (2000) and Machacek et
al. (2000). More complex is the picture which emerges from
hydrodynamical simulations of the IGM, where the mean gas
temperature appears to retain some memory of when and how it was
reionized (e.g. Schaye et al. 2000). Observational constraints
on $T_0(z)$ should also be taken into account (e.g. Shaye
et al. 1999, 2000; Bryan \& Machacek 2000; Ricotti et al. 2000;
McDonald et al. 2000).

In view of this variety of
assumptions and results it seems reasonable to look for solutions
of our equations assuming a general exponent $\alpha$ (although,
in practical applications we will focus on cases with $\alpha \leq
1$). Moreover, as we will see later, our model can be
straightforwardly extended to any redshift dependence of the mean
IGM temperature.

Solutions of the hydrodynamical equations for general values of
$\alpha$ have been obtained by Bi, B\"orner and Chu (1992). Many
authors (Peebles 1984; Soloveva \& Starobinskii 1985; Shapiro,
Giroux \& Babul 1994; Nusser 2000) have given solutions for the
case $\alpha=1$. In particular, Nusser (2000) has obtained the
linear IGM overdensity in the case $\alpha=1$, extending his
calculation to the case of non-negligible baryon fraction (i.e.
for $0\leq f_{\rm b} \leq 1$), i.e. accounting for the baryon
self-gravity. We will give here an extensive presentation of this
problem for general values of $\alpha$, both because we are going
to use the linear baryon overdensity in our nonlinear model for
the IGM dynamics, and because of the specific form taken by the
solutions for our set of initial conditions, which had not been
previously obtained in the literature. The particular case
$\alpha=1$ needs to be studied separately; this will be done in
the next subsection.

Let us start by writing the linearized continuity and Euler equations for
the baryonic component,
\be
{\dot \delta_{\rm b}}  =  - {\bf \nabla}\cdot{\bf u}_{\rm b} \;
\qquad;\qquad
{\dot {\bf u}}_{\rm b}
=-{3\over 2 a}\left({\bf u}_{\rm b}+
{\bf \nabla}\varphi +{2 \gamma k_B T_0 \over 3 H_0^2\Omega_{0{\rm m}}\mu m_p}
{\bf\nabla}\delta_{\rm b}\right) \;.
\ee

Combining these equations together, using the linear solutions for
the DM and our temperature-redshift relation we obtain, in Fourier
space,
\be
{\ddot \delta}_{\rm b}({\bf k},a)+{3\over 2 a} {\dot
\delta}_{\rm b}({\bf k},a)+{3\over 2 a^{\alpha+1}}
{k^2 \over {\tilde k}^2_J}
\delta_{\rm b}({\bf k},a)={3\over 2 a} \delta_{\rm DM}({\bf k}) \;.
\label{astrid}
\ee
The redshift-independent wavenumber ${\tilde k}_J$ is related to
the comoving Jeans wavenumber $k_J$ through
\be
k^2_J =  a^{\alpha-1}{\tilde k}_J^2 =
{3H_0^2\Omega_{0{\rm m}}\mu m_p (1+z) \over \gamma k_B T_0}\;.
\ee
Only for $\alpha=1$ the two wavenumbers coincide and the Jeans
length becomes a constant.

Equation (\ref{astrid}) will be solved with the initial, or, more
precisely, matching conditions at $a=a_i$,
\be
\delta_{\rm b}({\bf
k},a_i)=\delta_{\rm DM}({\bf k},a_i)
\quad\quad ; \quad\quad
{\dot \delta}_{\rm b}({\bf k},a_i)=
{\dot \delta}_{\rm DM}({\bf k},a_i) = {\delta_{DM}({\bf k},a_i)\over a_i}
= {\delta_{DM}({\bf k},a)\over a} \;,
\label{initial}
\ee
which are appropriate if the IGM undergoes sudden reionization
at $z_i$ (Nusser 2000).

\subsubsection{Case $T_0(z) \propto (1+z)$}

In the case $\alpha=1$, we try a solution of the type $\delta_{\rm
b}=A a^n+B a$, where $n$ and $B$ can be found by substitution back
into equation (\ref{astrid}). This gives
\be
\delta_{\rm b}({\bf
k},a)= a^{-1/4} \left[ {\tilde A}_1 a^{Q/4} + {\tilde A}_2
a^{-Q/4} \right] +{{\delta_{\rm DM}({\bf k},a) \over
(1+k^2/{{k_J}}^2)}} \;,
\label{29}
\ee
where $Q=\sqrt{1-24k^2/k_J^2}$ and ${\tilde A}_1$ and ${\tilde
A}_2$ are integration constants. From the latter expression we see
that on large scales the homogeneous part only contains decaying
modes; on smaller scales, instead, the homogeneous part is
characterized by an oscillatory behavior with decaying amplitude.
Asymptotically in time one recovers the well known solution (e.g.
Peebles 1993) $\delta_{\rm b}({\bf k},a) \approx \delta_{\rm
DM}({\bf k},a)/ (1+k^2/k_J^2)$. 

In spite of the absence of growing modes, the homogeneous part of
our solution does not necessarily become negligible at the time
scales relevant to our problem. It is therefore important, to
exactly evaluate the homogeneous part, by relating it to our
initial conditions above. For the two constants of integration we
find ${\tilde A_1} = {\tilde B_1} \delta_{\rm DM}({\bf k},a_i)$
and ${\tilde A_2} = {\tilde B_2} \delta_{\rm DM}({\bf k},a_i)$,
where
\be
{\tilde B_1}= {5+Q\over 2Q} \left({1\over 1+k^2/k_J^2}
\right) \left({k\over k_J}\right)^2 \;
\quad\quad ; \quad\quad
{\tilde B_2}= - {5-Q\over 2Q} \left({1\over 1+k^2/k_J^2}\right)
\left({k\over k_J}\right)^2 \;.
\label{tildeb}
\ee

The baryon peculiar velocity immediately follows from the linear continuity
equation. We obtain
\be
{\bf u}_{\rm b}=
\left[1+\left({k\over k_J}\right)^2\left({a\over a_i}\right)^{-5/4}
\left({5+Q\over 2Q} \left({a\over a_i}\right)^{Q/4}
- {5-Q\over 2Q}\left({a\over a_i}\right)^{-Q/4}
\right)\right]
{{\bf u}_{\rm DM}\over 1+k^2/k_J^2} \;.
\label{ufilter1}
\ee

\subsubsection{General case: $T_0(z) \propto (1+z)^\alpha$}

We will now look for the solution of eq. (\ref{astrid}), for arbitrary
values of $\alpha$, excluding the case $\alpha=1$.
In order to find the full solution to the inhomogeneous equation
(\ref{astrid}), we first notice that its homogeneous counterpart
is solved in terms of Bessel functions, namely,
$\delta_{\rm hom}({\bf k},a) =
y^{-{\tilde\alpha}} [ A_1 {\cal J}_{\tilde\alpha}(y) +
A_2 {\cal J}_{-\tilde\alpha}(y)]$,
where $\tilde\alpha = 1/[2(1-\alpha)]$ and we introduced the independent
variable $y \equiv \sqrt{24}{\tilde\alpha}(k/{\tilde k}_J)
a^{1/4{\tilde\alpha}}
= \sqrt{24} {\tilde\alpha} k/k_J$. The full solution of eq. (\ref{astrid})
for any $\alpha$, obtained by the standard Green's method, reads
\be
\delta_{\rm b}({\bf k},a)=y^{-{\tilde\alpha}}
\left[A_1{\cal J}_{\tilde\alpha}(y)+
A_2{\cal J}_{-\tilde\alpha}(y)\right]
+ \left[24 {\tilde\alpha}^2
y^{-5{\tilde\alpha}}
S_{5{{\tilde\alpha}}-1,{{\tilde\alpha}}}(y)
\right]\delta_{\rm DM}({\bf k}, a) \;,
\label{har}
\ee
where
$ S_{\mu,{\tilde \alpha}}(y) $ is the Lommel function.
%\be
%S_{\mu,{\tilde \alpha}}(y)={\pi\over 2}
%\left[{\cal J}_{-{\tilde \alpha}}(y)\int_0^y y^\mu
%{\cal J}_{{\tilde \alpha}}(y) dy-{\cal J}_{{\tilde
%\alpha}}(y)\int _0^y y^{\mu}
%{\cal J}_{-{\tilde \alpha}} (y) dy\right]
%\ee
%is the Lommel function.

By imposing the initial conditions
(\ref{initial}) on our solution, we obtain
$A_1 = B_1 \delta_{\rm DM}({\bf k},a_i)$ and
$A_2 = B_2 \delta_{\rm DM}({\bf k},a_i)$, where
\ba
B_1 & = &
-{\pi y_i^{1+{\tilde\alpha}}\over 2 {\rm sin}(\pi{\tilde\alpha})}
\left[\left(
1-24{\tilde\alpha}^2{S_{5{\tilde\alpha}-1,
{\tilde\alpha}}\over y_i^{5{\tilde\alpha}}}
\right) {\cal J}_{-{({\tilde\alpha}+1)}}
-{4{\tilde\alpha}\over y_i}
\biggl( 1-12{\tilde\alpha}(2{\tilde\alpha}-1)
y_i^{1-5{\tilde\alpha}}
S_{5{\tilde\alpha}-2,{\tilde\alpha}+1}
\biggr) {\cal J}_{-\tilde\alpha}
\right] \;,
\nonumber\\
B_2 & = &
-{\pi y_i^{1+{\tilde\alpha}}
\over 2{\rm sin}(\pi{\tilde\alpha})}
\left[
\left(1-24{\tilde\alpha}^2{S_{5{\tilde\alpha}-1,
{\tilde\alpha}}\over y_i^{5{\tilde\alpha}}}
\right) {\cal J}_{{\tilde\alpha}+1}
+ {4{\tilde\alpha}\over y_i}
\biggl( 1-12{\tilde\alpha} (2{\tilde\alpha}-1)
y_i^{1-5{\tilde\alpha}} S_{5{\tilde\alpha}-2,{\tilde\alpha}+1}
\biggr) {\cal J}_{\tilde\alpha}
\right] \;,
\label{generalb}
\ea
and the Bessel and Lommel functions are evaluated at $y=y_i=
\sqrt{24}{\tilde\alpha}(k/{\tilde k}_J)a_i^{1/4{\tilde\alpha}}$.

Once again, by replacing this expression in the continuity
equation, we obtain the linear peculiar velocity of the baryons,
\be
{\bf u}_{\rm b}({\bf k},a)= \left[
-{y_i^{4{\tilde\alpha}}\over 4{\tilde\alpha}}
\left(B_1 {\cal J}_{{\tilde\alpha}+1}-
B_2 {\cal J}_{-({\tilde\alpha}+1)}\right)
+ 12{\tilde\alpha}(2{\tilde\alpha}-1)
S_{5{\tilde\alpha}-2,{\tilde\alpha}+1}
\right] y^{1-5{\tilde\alpha}}
{\bf u}_{\rm DM}({\bf k}) \;.
\label{u90}
\ee
\subsection{IGM linear filter}

The results of the previous section indicate that one can always
express the linear baryon density field as a linear convolution of
the linear DM one. In Fourier space, one can therefore define an
IGM linear filter, $W_{\rm b}(k,a)$, such that
\be
\delta_{\rm
b}({\bf k},a) = W_{\rm b}(k,a)  \delta_{\rm DM}({\bf k}, a) \;.
\ee

For the case $\alpha=1$ we obtain
\be
W_{\rm b}(k,a) =
\left\{1+\left({k\over k_J}\right)^2
\left({a\over a_i}\right)^{-5/4} \hskip -0.1cm
\left[{5+Q\over 2Q}
\left({a\over a_i}\right)^{Q/4}
\hskip -0.1cm
-{5-Q\over 2Q}
\left({a\over a_i}\right)^{-Q/4}\right]\right\}
{1\over 1+k^2/k_J^2} \;,
\label{filt}
\ee
which corresponds to eq. (A7) of (Gnedin \& Hui 1998).
Let us look at the main features of the IGM filter, at a finite time after
reionization. It is immediate to check that on large scales
$W_{\rm b}$ tends to unity: on scales much larger than the baryon Jeans
length, the baryon distribution always traces the DM,
due to the negligible role of pressure gradients.
On scales much smaller than the Jeans length, instead, the baryon overdensity
undergoes rapid oscillations. As noticed by previous authors
(e.g. Gnedin \& Hui 1998; Nusser 2000), and contrary to naive expectations,
$W_{\rm b}(k,a)$ has no power-law asymptote on small scales.

In the general case $\alpha \neq 1$, instead we have
\ba
W_{\rm b}(k,a) & = &
24{\tilde\alpha}^2y^{-5{\tilde\alpha}}
S_{5{\tilde\alpha}-1,{\tilde\alpha}}(y)
\nonumber\\
& - & \left({y\over y_i}\right)^{-5{\tilde\alpha}}
{\pi y_i\over 2 {\rm sin}(\pi{\tilde\alpha})}
\left[ \left(
{\cal J}_{{\tilde\alpha}}(y){\cal J}_{-({\tilde\alpha}+1)}(y_i)+
{\cal J}_{-{\tilde\alpha}}(y){\cal J}_{{\tilde\alpha}+1}(y_i)
\right) \left(1-24{\tilde\alpha}^2
y_i^{-5{\tilde\alpha}}
S_{5{\tilde\alpha}-1,{\tilde\alpha}}(y_i)\right)
\right]
\nonumber\\
& - &
\left({y\over y_i}\right)^{-5{\tilde\alpha}}
{2\pi{\tilde\alpha}\over {\rm sin}(\pi{\tilde\alpha})}
\left[
\left(
{\cal J}_{-{\tilde\alpha}}(y){\cal J}_{{\tilde\alpha}}(y_i)-
{\cal J}_{{\tilde\alpha}}(y){\cal J}_{-{\tilde\alpha}}(y_i)
\right) \left(
1-12{\tilde\alpha}(2{\tilde\alpha}-1)
y_i^{1-5{\tilde\alpha}}
S_{5{\tilde\alpha}-2,{\tilde\alpha}+1}(y_i)
\right)
\right] \;.
\label{pra}
\ea
It is worth mentioning that for half-integer values of $\tilde \alpha$
the Bessel and the Lommel functions in the above equation can be written 
in terms of trigonometric functions.  

Analogous relations hold for the baryon peculiar velocity in terms of the
DM one [see eqs. (\ref{ufilter1}) and (\ref{u90}) above].
Also in this case, of course, $W_{\rm b} \to 1$ in the large-scale limit,
as it would be easy to see. The oscillating behavior 
of $\delta_{\rm b}$ on small scales
is in general more complicated and depends upon the value of $\alpha$.

Plots of the IGM linear filter for some values of $\alpha$ are
given in Figure 1, at various redshifts. Hydrogen reionization is
assumed to have occurred at a redshift $z_i=7$. Note that, on
scales smaller than the baryon Jeans length, acoustic 
oscillations, with decaying amplitude, persist even long
after reionization. Note also that different equations of state lead to a
different small-scale behaviour even if the IGM density is plotted 
vs. the scaled wavenumber $\kappa \equiv k/{\tilde k}_J$. 
This is because  eq. (\ref{astrid}), which governs the linear evolution of 
baryon density fluctuations, retains a dependence on the slope $\alpha$ in 
the last term on the LHS, which describes the effect of pressure gradients.

%%%%%%%%%%%%%%%%%%%%% FIGURE 1 %%%%%%%%%%%%%%%%%%%%%%%%%%%%%
%%%%%%%%%%%%%%%%%%%%% FIGURE 1 %%%%%%%%%%%%%%%%%%%%%%%%%%%%%
\begin{figure}
\hskip 1.85cm\epsfxsize= 9.7truecm{\epsfbox{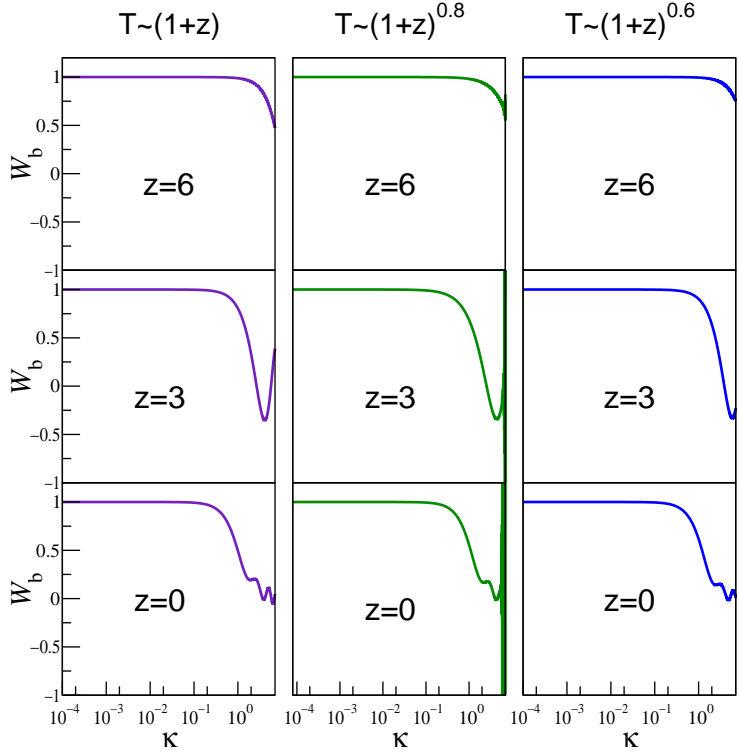}}
\caption{Plots of the ratio of the baryon
to dark matter density, 
$W_{\rm b} = \delta_{\rm b}/\delta_{\rm DM}$,
versus $\kappa \equiv k/{\tilde k}_J$ 
at different redshifts, for
various values of $\alpha$ in the mean IGM
temperature-redshift relation, 
$T_0(z) \propto (1+z)^\alpha$.
Hydrogen reionization is taken at $z_i=7$.}
\end{figure}
%%%%%%%%%%%%%%%%%%%%%% FIGURE 1 %%%%%%%%%%%%%%%%%%%%%%%%%%%%%
%%%%%%%%%%%%%%%%%%%%%% FIGURE 1 %%%%%%%%%%%%%%%%%%%%%%%%%%%%%

\section{Modelling the weakly nonlinear IGM dynamics}

With our choice of time variable, the Euler equation, both for the DM and the
baryons, takes the simple form
\be
{D {\bf u} \over D a} = - \nabla \eta \;,
\label{mockforce}
\ee
where the potential $\eta$ takes a different expression for the two fluids.
As the force acting on both fluids is conservative, in the
absence of initial vorticity the flow remains irrotational (actually,
this is only true prior to shell-crossing, for the DM component),
and we can express the peculiar velocity in terms of a velocity potential,
${\bf u} \equiv \nabla \Phi$.
The potential $\eta$ in the RHS of the Euler equation reads
\be
\eta_{\rm DM} = \frac{3}{2a} \left(\Phi_{\rm DM} + \varphi \right)
\qquad;\qquad
\eta_{\rm b} = \frac{3}{2a} \left[\Phi_{\rm b} + \varphi +
{1 \over (\gamma -1) a k_J^2} \left(1+
\delta_{\rm b}\right)^{\gamma-1}\right] \;,
\ee
for the DM and IGM components respectively.

The dynamical model we introduce here is largely inspired by the
Zel'dovich approximation (Zel'dovich 1970), and is based on
replacing the potential $\eta$, which should be consistently
calculated using the full set of equations for the two fluids, by
a mock external potential, obtained by evaluating its expression
within linear (Eulerian) perturbation theory. Indeed, this model
might be seen as only the first step of a more general, iterative
approximation scheme.

In order to evaluate to linear order the quantity $\eta$ one can
either compute to first order any single contribution or compute
its Laplacian by taking the divergence of the LHS of the
linearized Euler equation. This yields $\nabla^2 \eta\approx
- \nabla \cdot {\dot {\bf u}}$, and, using the linearized
continuity equation,
$
\nabla^2 \eta \approx {\ddot \delta} \;.
$

We thus conclude that, to first order in perturbation theory the
potential $\nabla^2 \eta$ is given by the second `time'-derivative
of the linear overdensity. This fact immediately implies that, to
first order, $\eta_{\rm DM}=0$ (neglecting the contribution 
from decaying modes), while
\be
\eta_{\rm b} (k,a) =
\left[a {\ddot W}_{\rm b}(k,a) + 2 {\dot  W}_{\rm b}(k,a) \right]
\varphi_0(k) \equiv {\cal E}(k,a) \varphi_0(k) \;.
\ee
In particular, for $\alpha=1$ only the homogeneous part of the
$\delta_{\rm b}$ solution contributes, and we obtain
\be
{\cal E}(k,a)= \left({a\over a_i}\right)^{-5/4}
\left[{(Q-1)(Q-5)\over 16a} \left({a\over a_i}\right)^{Q/4}
{\tilde B}_1 +{(Q+1)(Q+5)\over 16a} \left({a\over
a_i}\right)^{-Q/4} {\tilde B}_2\right] \;,
\ee
with $Q=\sqrt{1-24k^2/k_J^2}$ and the integration constants
${\tilde B}_1$ and ${\tilde B}_2$ as in eq. (\ref{tildeb}). For
$\alpha \neq 1$ we get
\ba
{\cal E}(k,a)&=&\hskip -0.1cm 
y_i^{4 {\tilde \alpha}} \left(\sqrt{24} {\tilde\alpha} {k\over
{\tilde k_J}}\right)^{4{\tilde\alpha}}
{{y^{1-9{\tilde\alpha}}}\over 16 {\tilde\alpha}^2} \biggl(
B_1\bigl[(4{\tilde\alpha}-2) {\cal J}_{{\tilde\alpha}+1}(y) +y
{\cal J}_{{\tilde\alpha}+2}(y)\bigr]
+ B_2\bigl[(2-4{\tilde\alpha}) {\cal J}_{-({\tilde\alpha}+1)}(y)
+ y{\cal J}_{-({\tilde\alpha}+2)}(y)\bigr] \biggr)
\nonumber\\
&+& \left(\sqrt{24} {\tilde\alpha} {k\over {\tilde
k_J}}\right)^{4{\tilde\alpha}}
y^{1-9{\tilde\alpha}}(12{\tilde\alpha}-6) \bigl[
2({\tilde\alpha}-1)y S_{5{\tilde\alpha}-3,{\tilde\alpha+2}}
+(1-2{\tilde\alpha}) S_{5{\tilde\alpha}-2,{\tilde\alpha}+1}
\bigr] \;,
\ea
where $B_1$ and $B_2$ are given by eq. (\ref{generalb}).
Plots of the function ${\cal E}(k,a)$ are given in Figure 2,
at different redshifts after reionization, for various values of
$\alpha$.

%%%%%%%%%%%%%%%%%%%%% FIGURE 2 %%%%%%%%%%%%%%%%%%%%%%%%%%%%%%%%
%%%%%%%%%%%%%%%%%%%%% FIGURE 2 %%%%%%%%%%%%%%%%%%%%%%%%%%%%%%%%
\begin{figure}
\hskip 1.845cm\epsfxsize= 9.7truecm{\epsfbox{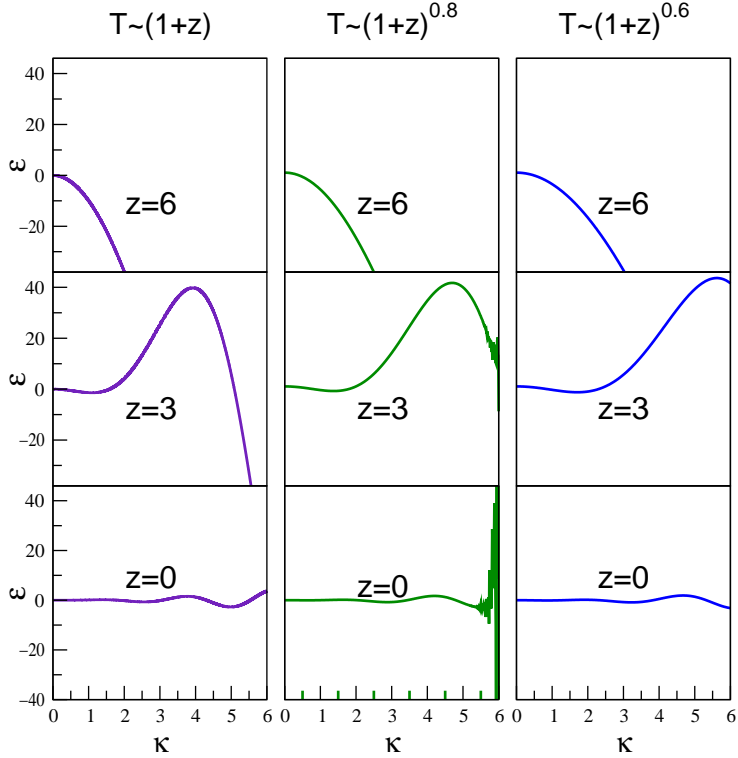}}
\caption{Plots of the ratio ${\cal E} = \eta/\varphi$
versus $\kappa \equiv k/{\tilde k}_J$ at different redshifts,
for various values of $\alpha$. The reionization redshift is
$z_i=7$.}
\end{figure}
%%%%%%%%%%%%%%%%%%%%% FIGURE 2 %%%%%%%%%%%%%%%%%%%%%%%%%%%%%%%%
%%%%%%%%%%%%%%%%%%%%% FIGURE 2 %%%%%%%%%%%%%%%%%%%%%%%%%%%%%%%%

Therefore, our model is described by the two Euler equations:
\be
{D {\bf u}_{\rm DM} \over D a} = 0 \;, \label{effeulerdm}
\ee
for the DM component, and
\be {D {\bf u}_{\rm b} \over D a} = -
\nabla \eta_{\rm b}\;,
\label{effeulerigm}
\ee
for the baryons.

The solution of the above DM equation of motion is well known: it
corresponds to the Zel'dovich approximation, according to which
mass elements move along straight lines, with constant `velocity'
impressed by local linear fluctuations of the gravitational
force at their initial {\it Lagrangian} location $\bf q$: ${\bf
u}_{\rm DM}({\bf x}({\bf q},a),a) = - \nabla_{\bf q}
\varphi_0({\bf q})$. We therefore have
\be
{\bf x}_{\rm DM}({\bf
q},a) = {\bf q} - a \nabla_{\bf q} \varphi_0({\bf q}) \;.
\label{dmtraj}
\ee

The trajectories of the IGM fluid elements are instead more
complex, as our equation implies a non-zero force acting on them,
\be
{\bf x}_{\rm b}({\bf q},a) = {\bf q} - a \nabla_{\bf q}
\varphi_0({\bf q}) - \int_0^a d\tau \int_0^\tau d \tau^\prime
~\nabla_{\bf x} \eta({\bf x}({\bf q},\tau^\prime),\tau^\prime) \;,
\label{igmtraj}
\ee
where we have formally extended the time integration from $0$ to
$a$, as, before reionization, when the Jeans length is negligible,
$\eta_{\rm b}({\bf x},a<a_i)=0$.

It will also prove convenient in the following to transform the
baryon Euler equation (\ref{effeulerigm}) into the following
Bernoulli, or Hamilton-Jacobi, equation for the velocity
potential:
\be
{\partial \Phi_{\rm b}\over\partial a}+ {1\over
2}(\nabla\Phi_{\rm b})^2 =  - \eta_{\rm b} \;.
\label{bernoulli1}
\ee

The extreme simplicity of our scheme is shown by the fact that the
only information needed to evolve the baryon distribution in the
weakly nonlinear regime is the IGM filter $W_{\rm b}$. It is
immediate to realize that, because of our derivation of $\eta_{\rm
b}$, its expression, $\nabla^2 \eta \approx {\ddot \delta}$, in terms of the
linear baryon overdensity is fully general, i.e. it would apply to
general reionization histories [i.e. to general $T_0(z)$
relations], general baryon equations of state and to the case in
which the baryon self-gravity is properly taken into account (i.e. 
the general case $0 < f_{\rm b} \leq 1$). Note that the IGM linear
filter is simply the ratio of the baryon to DM transfer function
at the given time $W_{\rm b}(k,a) = T_{\rm b}(k,a) /T_{\rm
DM}(k,a)$ (provided the reionization process has been taken into
account in evaluating the baryon transfer function $T_{\rm b}$).

One might think that there is some degree of arbitrariness in the
particular choice we made of which terms to linearize [those on
the RHS of eq. (\ref{mockforce})] and which ones to treat exactly
(those on its LHS). The strongest motivation for such a choice is
its analogy with the procedure leading to the Zel'dovich
approximation for the DM component. In this sense, the choice is
unique; as we will see in the next subsection, in fact, the
external force $\nabla \eta_{\rm b}$ obtained above and its effect
on the IGM trajectories are closely connected to the results of
first-order Lagrangian perturbation theory: no other choice would
have provided such a connection. It should also be stressed that
other successful approximation schemes for the evolution of the DM
component, such as the `frozen flow' (Matarrese et al. 1992) and
`frozen potential' (Brainerd, Scherrer \& Villumsen 1993; Bagla \&
Padmanabhan 1994) ones, are indeed based on the same choice. The
physical reason which makes a linear theory evaluation of $\eta$
reasonably accurate is that this quantity, for both fluids,
contains terms that receive their dominant contribution from small
wavenumbers. This was indeed the original motivation which led
Zel'dovich to obtain his celebrated algorithm, although in modern
language the Zel'dovich approximation is more commonly explained
within the first-order Lagrangian approximation (e.g. Sahni \& Coles
1995 and references therein). The link between these two alternative
derivations is discussed in the next subsection.

\subsection{Lagrangian approximation to the baryon trajectories}

The trajectory of any mass element can be written in the general
form \be {\bf x}({\bf q},a) = {\bf q} + {\bf S}({\bf q},a) \;, \ee
where ${\bf S}$ is the `displacement vector'. As far as the
evolution of the fluid is far from the strongly nonlinear regime,
the displacement vector can be considered to be small, i.e. the
Eulerian and Lagrangian positions of each particle are never too
far apart. This consideration is at the basis of Lagrangian
approximation methods. The basics of the first-order Lagrangian
scheme applied to our two-component 
fluid are reported in Appendix A. Applying these ideas to our
external force, we obtain 
\be \nabla_{\bf x} \eta_{\rm b}({\bf
x}({\bf q},a), a) = \nabla_{\bf q} \eta_{\rm b}({\bf q}, a) +
{\cal O}\left({\bf S}^2\right) \;.
\label{approxeta} 
\ee

To first order in the displacement vector we can therefore replace
the Eulerian force $\nabla_{\bf x} \eta_{\rm b}({\bf x}, a)$ by
its Lagrangian counterpart $\nabla_{\bf q} \eta_{\rm b}({\bf q},
a)$ in the baryon trajectories, which leads to the much simpler
form
\be
{\bf x}_{\rm b}({\bf q},a) = {\bf q} - a \nabla_{\bf q}
\psi_{\rm b}({\bf q},a) \;, 
\label{newtraj} 
\ee 
where the `baryon potential' $\psi_{\rm b}$ is defined by 
\be 
\nabla^2_{\bf
q}\psi_{\rm b}({\bf q},a) = {\delta_{\rm b}({\bf q}, a) \over a}
\label{barpot}
\ee
and is related to the peculiar gravitational potential by the
Fourier-space expression $\psi_{\rm b}({\bf k},a) = W_{\rm b}(k,a)
\varphi_0({\bf k})$.

Moreover, as shown in Appendix A, the trajectories described by
eq. (\ref{newtraj}) represent the result of first-order Lagrangian
perturbation theory applied to our baryon gas. Similar results
have been recently obtained by Adler \& Buchert (1999) for the
case of a single self-gravitating collisional fluid (i.e. for
$f_{\rm b}=1$). There is an important difference between these
trajectories and the DM ones. According to the Zel'dovich
approximation, DM particles move along straight lines with
constant velocity, whereas the baryons are generally accelerated
along curved paths; this is due to the non-zero force, resulting
from the composition of three terms: the local Hubble drag, the
local gravitational force caused by the dominant DM component and
the gradient of the gas pressure.

The baryon peculiar velocity which follows from this first-order
Lagrangian approximation is
\be
{\bf u}_{\rm b}({\bf x}({\bf
q},a),a) = - \nabla_{\bf q} \left(a {\dot \psi}_{\rm b}({\bf q},a)
+ \psi_{\rm b}({\bf q},a) \right) \;,
\label{newvel}
\ee
which, unlike the DM one, deviates from the initial velocity ${\bf
u}_{{\rm b}0}({\bf q}) \equiv {\bf u}_{\rm b}({\bf x}({\bf
q},0),0) = - \nabla_{\bf q} \varphi_0({\bf q})$.

Let us mention an important property of the approximate expression
for the velocity field that we obtained using first-order
Lagrangian perturbation theory: unlike what happens in the DM
case, the vector ${\bf u}_{\rm b}$ of eq. (\ref{newvel}) is
irrotational in Eulerian space only if we restrict its validity to
first-order in the displacement vector; a non-zero velocity curl
component, $\nabla_{\bf x} \times {\bf u}_{\rm b} \neq 0$, in fact
arises beyond this limit. This feature is not expected to imply
serious problems as long as the system has not evolved too deeply
into the nonlinear regime, i.e. until the baryon overdensity has
not reached values $\gg 1$. This is an unphysical feature deriving
from the extrapolation of the first-order results to a regime
where higher-order terms should be taken into account.
A similar feature appears in the Lagrangian perturbation approach
to the DM dynamics, when both the growing and decaying solutions are
included (e.g. Buchert 1992). No problems
of this type, however, occur when the Eulerian scheme above is
adopted, i.e. when eq. (\ref{effeulerigm}) and its solution
(\ref{igmtraj}) are assumed. For this reason we will consider our
`Eulerian model' of eqs. (\ref{effeulerigm}) and (\ref{igmtraj})
as the correct one and the Lagrangian scheme leading to eq.
(\ref{newtraj}) as being essentially a shortcut to get
approximated baryon trajectories.

The slight discrepancy between the Eulerian and Lagrangian schemes
used to derive the external force $\nabla \eta$ acting on the
baryons is a peculiarity of the collisional case. In the DM case,
these two techniques -- linearizing the RHS of eq.
(\ref{mockforce}) in Eulerian space, and expanding to first order
in Lagrangian perturbation theory -- lead to identical results.

Approximation schemes for the low-density IGM dynamics which are
closely related to our Lagrangian treatment have been studied by
Reisenegger \& Miralda-Escud\'e (1995), Gnedin \& Hui (1996), and
Hui, Gnedin \& Zhang (1997). There are, however, important
differences that we would like to point out: {\it i)} in our model
the gas trajectories, and thus the resulting IGM spatial
clustering depend on the specific ionization history (different
values of $\alpha$ in the simplest case); {\it ii)} our model is
by definition able to {\it exactly} reproduce the behavior of the
baryon component in the linear regime. In practice, while previous
models adopt an IGM filter which is best-fitted to
simulations, ours directly derives from baryon dynamics.

\subsubsection{Numerical simulations of the IGM distribution in the
laminar regime}

In order to display the effect of the IGM filter on baryon
trajectories, we have produced a set of numerical simulations
based on the Lagrangian scheme. $256^3$ particles were laid down
on a uniform cubic grid and moved according to eq.
(\ref{newtraj}). Realizations of the peculiar gravitational
potential were obtained in Fourier-space according to standard
algorithms. The model shown in Figure 3 is a spatially flat,
vacuum dominated cold dark matter (CDM) model, with $h=0.65$
(where $h$ is the Hubble constant, $H_0$, in units of $100$ km
s$^{-1}$ Mpc$^{-1}$), a cosmological constant with present-day
density parameter $\Omega_{0\Lambda} = 0.7$, CDM with $\Omega_{0
{\rm CDM}} = 0.26$ and baryons with the remaining $\Omega_{0 {\rm
b}}=0.04$. A Zel'dovich primordial power-spectrum of adiabatic
perturbations is assumed, with the standard normalization of
$\Lambda$CDM models\footnote{To correct for the small inaccuracy
introduced by the use of $a$, instead of the DM growth-factor, as
time variable in our treatment, we apply a correction factor
$(1+z)D_+(z)$ to the $\sigma_8$ normalization, where $D_+(z)$ is
the linear growth-factor of DM density fluctuations, normalized to
unity at $z=0$.}, $\sigma_8=0.99$ (Viana \& Liddle 1999),
where $\sigma_8$ is the {\it rms} mass fluctuation in a
sharp-edged sphere of radius $8~h^{-1}$ Mpc; we adopt the Bardeen
et al. (1986) CDM transfer function, corrected to account for the
baryon contribution, as in (Sugiyama 1995).

The box-size is $10.24$ Mpc. Three different IGM thermal histories
are considered: in all cases we assume that reionization occurred
at $z_{i} =7$, but we assume that the mean IGM temperature evolves
according to different power-laws afterwards, $T_0(z)\propto
(1+z)^\alpha$, with $\alpha=1.0,~0.8,~0.6$. Figure 3 shows
a thin ($0.02$ Mpc) slice of our simulation box at $z=3$.
The corresponding DM distribution is also shown for comparison.
Note that different values of $\alpha$ lead to a different
small-scale distribution of the IGM, because of {\it i)} the
different time dependence of the Jeans scale and {\it ii)} the
different trajectories followed by the baryons after reionization.
The comoving Jeans wavenumber $k_J$ has been set to the same
value ($7$ Mpc$^{-1}$) in all our models, at redshift $z=3$.

%%%%%%%%%%%%%%%%%%% FIGURE  3 %%%%%%%%%%%%%%%%%%%%%%%%%%%%%%%
%%%%%%%%%%%%%%%%%%% FIGURE  3 %%%%%%%%%%%%%%%%%%%%%%%%%%%%%%%
\begin{figure}
\begin{center}
\hskip -0.2cm\epsfxsize= 10truecm{\epsfbox{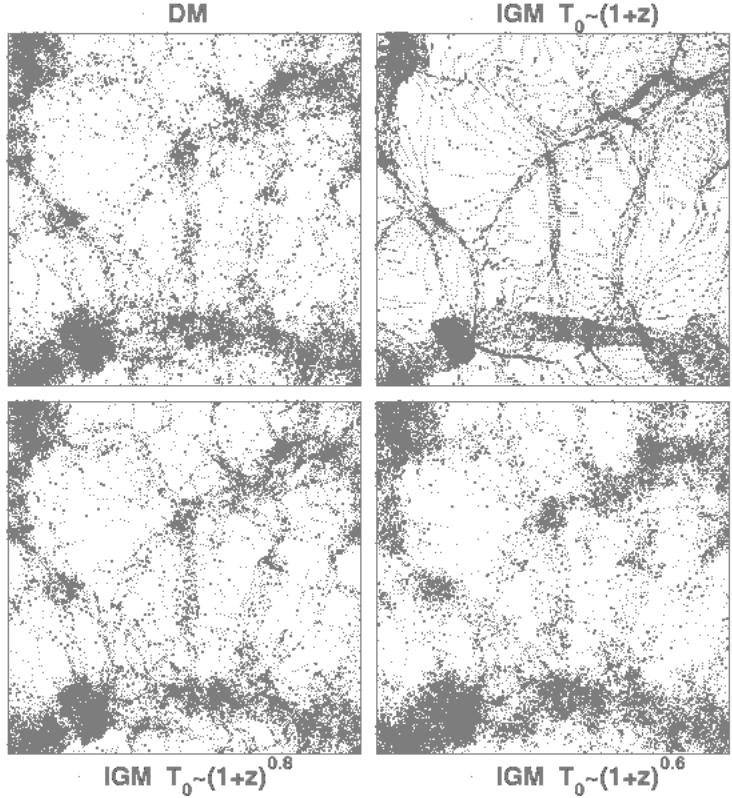}}
\end{center}
\caption{The spatial distribution of the dark matter and IGM, with 
different temperature-redshift relations, are shown at redshift $z=3$. 
All models have the same Jeans length, $k_J=7 {\rm Mpc}^{-1}$, 
at this redshift.}
\end{figure}
%%%%%%%%%%%%%%%%%%%%%% FIGURE 3 %%%%%%%%%%%%%%%%%%%%%%%%%%%%%
%%%%%%%%%%%%%%%%%%%%%% FIGURE 3 %%%%%%%%%%%%%%%%%%%%%%%%%%%%%

%%%%%%%%%%%%%%%%%%% FIGURE  4 %%%%%%%%%%%%%%%%%%%%%%%%%%%%%%%
%%%%%%%%%%%%%%%%%%% FIGURE  4 %%%%%%%%%%%%%%%%%%%%%%%%%%%%%%%
%\begin{figure}
%\begin{center}
%\hskip -0.2cm\epsfxsize= 16truecm{\epsfbox{Skj3z3.ps}}
%\end{center}
%\caption{The spatial distribution of the dark matter and IGM, with 
%different temperature-redshift relations, are shown at redshift $z=3$. 
%All models have the same Jeans length, $k_J=3 {\rm Mpc}^{-1}$, 
%at this redshift.}
%\end{figure}
%%%%%%%%%%%%%%%%%%%%%% FIGURE 4 %%%%%%%%%%%%%%%%%%%%%%%%%%%%%
%%%%%%%%%%%%%%%%%%%%%% FIGURE 4 %%%%%%%%%%%%%%%%%%%%%%%%%%%%%

\subsubsection{The IGM density field in the laminar regime}

Once particle trajectories are known, it is immediate to obtain
the density field, using mass conservation, $1+ \delta_{\rm
b}({\bf x},a) = |\!|\partial {\bf x}_{\rm b}({\bf q},a) / \partial
{\bf q}|\!|^{-1}$. If we adopt the Lagrangian expression of eq.
(\ref{newtraj}), we get
\be
1+ \delta_{\rm b}({\bf x}({\bf
q},a),a) = \left({\rm det}\left[\delta_{ij} -a {\partial^2
\psi_{\rm b}({\bf q},a) \over
\partial q_i \partial q_j} \right]\right)^{-1} \;.
\ee

The {\it strain tensor} $\partial^2 \psi_{\rm b}/\partial q_i
\partial q_j$ can be locally diagonalized along principal axes,
whose direction will generally depend upon time, unlike the DM
case. Calling $\lambda_{\rm i}$ the corresponding eigenvalues, we
write
\be
1+ \delta({\bf x}({\bf q},a),a) = \prod_{i=1}^3
\left[1-a \lambda_i({\bf q},a) \right]^{-1} \;,
\label{eigenvalues}
\ee
which shows that a caustic singularity will form at the finite time
$a_{\rm sc}({\bf q})=1/\lambda_\star({\bf q},a_{\rm sc})$, where
$\lambda_\star({\bf q},a_{\rm sc})$ is the largest eigenvalue of the
time-dependent strain tensor.
The time dependence of the eigenvalues, which on small scales
becomes oscillatory, implies that the shell-crossing condition can
even be met more than once by a given mass element along the same 
principal axis.

The extrapolation of
our Lagrangian approximation beyond the formation of the first
pancakes, leads to an artificial diffusion of these structures. 
This problem becomes more and more severe at low redshift, making 
any simplified description of the IGM -- and of the Ly$\alpha$ forest -- 
based on the Lagrangian trajectories quite unreliable. 
The stochastic adhesion model presented in the next section aims
precisely at overcoming this problem. 

%%%%%%%%%%%%%%%%%%%%%% FIGURE 5 %%%%%%%%%%%%%%%%%%%%%%%%%%%%%
%%%%%%%%%%%%%%%%%%%%%% FIGURE 5 %%%%%%%%%%%%%%%%%%%%%%%%%%%%%
%\begin{figure}
%\begin{center}
%\epsfxsize= 16truecm{\epsfbox{shellcrossing.ps}}
%\end{center}
%\caption{
%Orbit-crossing at $z=1$ is shown for the DM and different
%models of the IGM mean temperature evolution. The Jeans length is 
%taken to be $7$ Mpc $^{-1}$ at $z=3$, as in Figure 3.}
%\end{figure}
%%%%%%%%%%%%%%%%%%%%%% FIGURE 5 %%%%%%%%%%%%%%%%%%%%%%%%%%%%%
%%%%%%%%%%%%%%%%%%%%%% FIGURE 5 %%%%%%%%%%%%%%%%%%%%%%%%%%%%%

One might use eq. (\ref{eigenvalues}) for the density field to
obtain an analytical expression for the probability distribution
function (PDF) of the IGM density field, by a simple extension of
the traditional Doroshkevich (1970) formalism (e.g. Kofman et al.
1994). Such a technique has been applied by Reisenegger \&
Miralda-Escud\'e (1995), who adopted a smoothed Zel'dovich
approximation for the IGM evolution. Our model, however, allows to
get a more refined description in which the PDF is sensitive to
the IGM thermal history. Similarly, one might also obtain
statistical information on the Ly$\alpha$ forest in quasar
absorption spectra, e.g. in terms of the column density
distribution function of the lines (e.g. Hui, Gnedin \& Zhang
1997); this too would display a dependence on the assumed
$T_0(z)$, which might be tested against observational data. These,
and related applications of our Lagrangian algorithm will be
presented elsewhere.

The emergence of shell-crossing singularities in the velocity
pattern of our collisional fluid should be understood as an
artifact of the linearized treatment of pressure gradients in the
Euler equation. 
This feature can be also seen as follows. If we
evaluate the force $\eta$ on large scales, to lowest order in
$k/k_J$ we find, for $\alpha=1$,
\be
{\cal E}(k,a) \approx -
{3 \over 2 a} \left({a \over a_i}\right)^{-1/2} \left({k \over k_J}
\right)^2
\;,
\label{leadingforce1}
\ee
which comes only from the homogeneous part of eq. (\ref{29}), and
\be
{\cal E}(k,a) \approx - {6 \over a}
{(1- \alpha) \over (2 \alpha -1)(5 - 2 \alpha)} \left({k \over k_J}
\right)^2 \;,
\label{leadingforcenot1}
\ee
for $\alpha \neq 1$, from the inhomogeneous term of eq. (\ref{har}).
In both cases the result can be expressed in the form
$\nabla \eta \approx - \nu_{\rm eff}(a) \nabla^2 {\bf u}_0$,
where $\nu_{\rm eff}>0$, for $0.5<\alpha \leq 1$. This shows that,
as long as the velocity is in the linear regime, the effect of the
linearized pressure gradient is similar to that of an effective
kinematical viscosity\footnote{A related point has been recently made
by Buchert \& Dom\'inguez (1998).}, with coefficient $\nu_{\rm eff}$, which
smooths the velocity field. As soon as the system enters the
mildly non-linear regime, however, ${\bf u}$ deviates from its
initial value and this smoothing is no more effective in stopping
the formation of caustic singularities in the densest regions.

Two processes will actually prevent the formation of multi-streams
in our physical system: the first is due to the binding action of gravity,
which tends to keep pancakes and filaments thin, an effect which is
experienced both by the DM and IGM components;
the second is the nonlinear action of the gas pressure, which is
instead characteristic of the baryonic component. Because of the
difficulty to deal analytically with the gas pressure beyond any
perturbative approximation, we will find convenient in the next
section to introduce an artificial viscosity term in the baryon
Euler equation, whose effect is to smear these shell-crossing
singularities of the velocity field.

\section{The stochastic adhesion approximation}

The model for the formation of structures in the baryon
distribution discussed in the last section is based on a suitable
approximation for the force exerted on fluid elements by gravity
and by the surrounding fluid patches through pressure gradients.
The latter are most important on small scales where their effect
is to smooth the gas fluctuations relative to the DM ones, thus
preventing the occurrence of shell-crossing singularities in the
collisional component; nonlinear effects here manifest themselves
through the formation of shock-waves. In our scheme, just like in
all similar schemes [e.g. the modified Zel'dovich scheme used by
Gnedin \& Hui (1996) and Hui, Gnedin \& Zhang (1997)], the gas
pressure is only included through a linear approximation. Because
of this fact, when the system enters the nonlinear regime, i.e.
when two particles come very close in space, shell-crossing can
affect their evolution leading to the subsequent occurrence of
multi-streams. 
In order to prevent this unphysical phenomenon we
will adopt the same technique which proved so successful in the DM
case to extend the Zel'dovich treatment beyond its actual range of
validity: we add a kinematical viscosity term to our approximate
Euler equation. This method is at the basis of the adhesion
approximation for the formation of large-scale structure in the
collisionless case (Gurbatov et al. 1985, 1989; Kofman \&
Shandarin 1988). The adhesion model is based on the
three-dimensional generalization of Burgers' equation of strong
turbulence. It is the simplest equation which describes the
formation and subsequent merging of shocks. According to the
adhesion model, DM particles move according to the Zel'dovich
approximation until they fall into pancakes, when, owing to the
viscous force, they stick together. Next, pancakes drain
into filaments, and filaments into clumps. The thickness of
these structures is monitored by the value of the kinematical
viscosity coefficient $\nu$, and becomes infinitely thin as
$\nu \to 0$.

We assume that the equation of motion which governs the IGM
dynamics is (from now on we will avoid the subscript `b' on baryon
quantities, where unnecessary)
\be
{D {\bf u} \over D a} = \nu
\nabla^2 {\bf u} - \nabla \eta \;,
\label{stochadh}
\ee
where the kinematical viscosity coefficient $\nu$ is here assumed
to be small, but non-zero. In our collisional case, moreover,
there can be an extra reason to add such a term: the gas is
effectively experiencing some shear viscosity, although on scales
much smaller than those under consideration.
The physical viscosity term should actually depend upon space and
time, through the inverse of the local gas density. One might
however argue that, in the physically relevant limit, where only a
tiny viscous force is present, even a constant $\nu$ will produce
the correct qualitative trend.
It might be interesting to mention an alternative interpretation
of the viscosity term in self-gravitating systems. According to
Dom\'inguez (2000), a term which resembles the one for kinematical
viscosity is the unavoidable consequence of the {\it coarse-graining}
process inherent in a hydrodynamical description.

The equation above is known as the {\it forced Burgers equation}
of nonlinear 
diffusion (e.g. Barab\'asi \& Stanley 1995; Frisch \& Bec 2000).
Its possible application to the cosmological structure formation
problem was suggested long ago by Zel'dovich et al. (1985, 1987) and
recently studied in greater detail by Jones (1996, 1999). There are a
number of differences between our approach and the one by Jones,
which lead to a different dynamics of the IGM. First, we used a
different time variable (the scale-factor $a$ instead of the
cosmic time $t$) to define the baryon velocity field and its
acceleration, thus leading to the form of eq. (\ref{mockforce}),
where the force on the RHS was replaced by its linearized
expression. Second, we do include some effect of the baryonic
pressure in our external random potential $\eta$, which is instead
absent in Jones' treatment, where the role of $\eta$ is played 
by the linear gravitational potential generated by the dark matter 
component. Thus, our scheme, unlike the one by Jones, is able to exactly
reproduce the evolution of the baryons at the linear level.
Finally, our random potential has a non-trivial time dependence,
unlike the one assumed by Jones.

The forced Burgers equation can be transformed into a
Bernoulli-like equation for the velocity potential, namely
\be
{\partial \Phi\over\partial a}+ {1\over 2}(\nabla\Phi)^2
=\nu{\nabla}^2\Phi - \eta \;.
\label{bernoulli2}
\ee

The problem is fully specified once the initial conditions for the
velocity potential and the statistics of the noise are given. Our
initial velocity potential is $\Phi_0({\bf q}) = - \varphi_0({\bf
q})$. The statistics of the noise follows directly from that of
the linear gravitational potential, which we assume to be a
Gaussian random field. Then, also $\eta$ is a Gaussian process
with zero mean and auto-correlation function
\be
\langle \eta({\bf
x}, a) \eta({\bf x} + {\bf r}, a^\prime) \rangle = {1 \over
2\pi^2} \int_0^\infty {dk \over k^2} ~{\cal E}(k,a) {\cal
E}(k,a^\prime) P(k) j_0(kr)\;,
\ee
where the average is over the $\varphi_0$ ensemble, $P(k)$ is the linear
power-spectrum of DM density fluctuations, extrapolated to the
present time, and $j_0$ is the spherical Bessel function of order
zero. According to this formula, $\eta$ is a stochastic process
with the following properties:

\noindent {\it i)} It is a colored (i.e. non-white) Gaussian
random process both in space and time.

\noindent {\it ii)} It is stationary (i.e. homogeneous and
isotropic) in space but not in time (as its auto-correlation
function does depend on $|{\bf x}-{\bf x}'|$ only, but not on $|a
- a'|$: this is typical of the cosmological case.

\noindent {\it iii)} On small scales, $\eta$ oscillates rapidly in
time for all $\alpha\leq 1$, with a period which generally
depends on the wavenumber.

\noindent {\it iv)} The non-separability of the space and time
dependence of its auto-correlation function implies that $\eta$
behaves as a stochastic process both in space and time. More
precisely, the time evolution of the noise in each given spatial
point cannot be predicted on the basis of local initial conditions
only, as it depends on the realization of the Gaussian field
$\varphi_0$ over the whole Lagrangian space, through the
time evolution of the IGM filter. \\

Let us finally stress an important peculiarity of our cosmological
application of the forced Burgers equation: the same stochastic
process $\varphi_0$ which underlies the external random potential
$\eta = {\cal E} \ast \varphi_0$ also provides the initial condition
for the velocity potential, $\Phi_0 = - \varphi_0$.

A plot of the function ${\cal E}(k,a)$ vs. redshift is given in
Figure 4 for various values of $\alpha$, to display the
time-dependence of the random force and, in particular, its oscillatory 
character on small scales. In Figure 5 the 
dimensionless power-spectra $\Delta^2(k)$ of two terms contributing 
to the evolution of the velocity potential are shown: 
that of the random potential $\eta$, 
$\Delta^2_\eta(k,z) \equiv (1/2\pi^2) k^{-1} {\cal E}^2(k,z) P(k)$, and 
that of the linear velocity potential $\Phi_0$, multiplied by a 
factor $1/a=1+z$ to give an estimate of the `deterministic' term in 
eq. (\ref{bernoulli2}), namely
$\Delta^2_{(1+z)\Phi_0}(k,z) \equiv (1/2\pi^2) (1+z)^2 k^{-1} P(k)$.
Note that the two contributions become of the same order  
at $k \sim k_J$ at the relevant redshifts; 
on smaller scales the random potential dominates the IGM dynamics. 

%%%%%%%%%%%%%%%%%%%%% FIGURE 6-->4 %%%%%%%%%%%%%%%%%%%%%%%%%%%%%
%%%%%%%%%%%%%%%%%%%%% FIGURE 6-->4 %%%%%%%%%%%%%%%%%%%%%%%%%%%%%
\begin{figure}
\begin{center}
\hskip -1cm\epsfxsize= 12truecm{\epsfbox{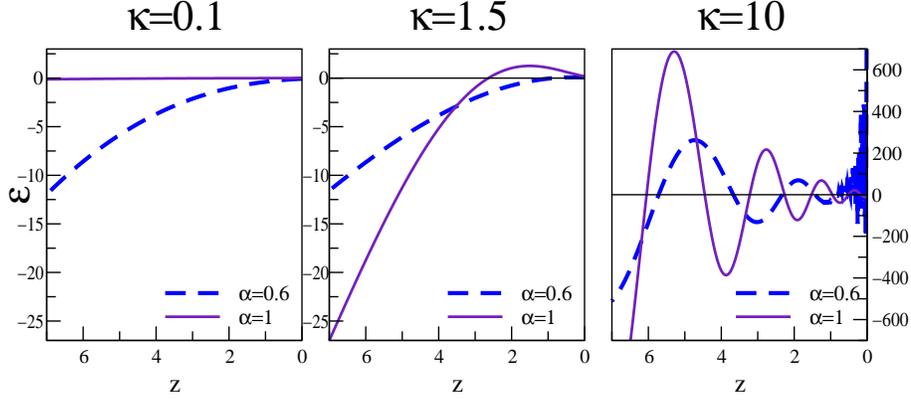}}
\end{center}
\caption{The evolution of the ratio ${\cal E} = \eta/\varphi_0$, for 
various models of the IGM mean temperature evolution.}
\end{figure}
%%%%%%%%%%%%%%%%%%%%%% FIGURE 6-->4 %%%%%%%%%%%%%%%%%%%%%%%%%%%%%
%%%%%%%%%%%%%%%%%%%%%% FIGURE 6-->4 %%%%%%%%%%%%%%%%%%%%%%%%%%%%%
%
%%%%%%%%%%%%%%%%%%%%% FIGURE 7-->5 %%%%%%%%%%%%%%%%%%%%%%%%%%%%%
%%%%%%%%%%%%%%%%%%%%% FIGURE 7-->5 %%%%%%%%%%%%%%%%%%%%%%%%%%%%%
\begin{figure}
\hskip 1.6cm\epsfxsize= 12truecm{\epsfbox{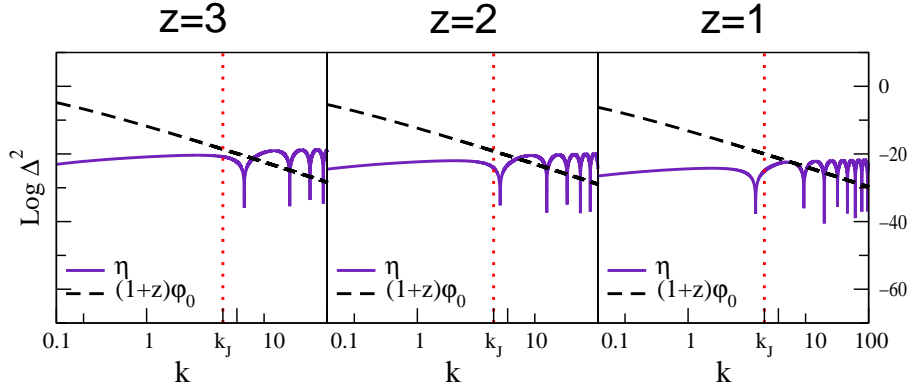}}

\caption{The dimensionless power-spectra of the stochastic and linear 
deterministic terms in the evolution of the IGM velocity potential 
are shown for $\alpha=1$ at $z=3$ (left panel), $z=2$ (middle panel)
and $z=1$ (right panel). The normalization of the linear density 
power-spectrum, $P(k)$, is chosen arbitrarily.
}
\end{figure}
%%%%%%%%%%%%%%%%%%%%%% FIGURE 7-->5 %%%%%%%%%%%%%%%%%%%%%%%%%%%%%
%%%%%%%%%%%%%%%%%%%%%% FIGURE 7-->5 %%%%%%%%%%%%%%%%%%%%%%%%%%%%%

The above evolution equation for $\Phi$, in cases when the random
potential $\eta$ is white-noise in time, has been extensively
studied in condensed matter physics, where it became popular as
the Kardar-Parisi-Zhang (KPZ) equation (Kardar, Parisi \& Zhang
1986). The KPZ equation describes the growth of an interface under
random particle deposition. Here the viscosity coefficient $\nu$
plays the role of temperature and the velocity potential $\Phi$
is interpreted as the height above an initially flat surface,
which is driven by the random noise and gradually becomes rough.

\subsection{Solving the random heat equation}

Equation (\ref{bernoulli2}) can be easily related to a linear,
partial-differential equation, by means of the nonlinear
Hopf-Cole transformation $\Phi= - 2\nu {\rm ln}\, {\cal
U}$ (e.g. Burgers 1974), which leads to the linear diffusion
equation with a multiplicative potential, also called `random
heat' equation
\be
{\partial {\cal U}({\bf x},a)\over\partial a}= \nu
\nabla^2{\cal U}({\bf x},a) +{\eta({\bf x},a)\over 2\nu}{\cal
U}({\bf x},a) \;.
\label{fa1}
\ee

Starting from the solution of the latter equation for the Hopf-Cole
transformed velocity potential ${\cal U}$ [which has been dubbed
`expotential' by Weinberg \& Gunn (1990a)] one can easily find the
velocity by transforming everything back
\be {\bf u} = -
{2\nu}{\nabla {\cal U}\over {\cal U}}
\label{reverse}
\ee

One first obtains an expression for the `transition kernel'
${\cal K}({\bf x},a |{\bf q},0)$, representing the particular solution
obtained from a Dirac delta function, $\delta({\bf x} - {\bf q})$, at the
initial time $a = 0$.
This has a formal solution given in terms of the (Euclidean) Feynman-Kac
path-integral
\footnote{The solution of eq. (\ref{fa1}) in the absence of a potential,
i.e. for the `free' adhesion approximation,
is reviewed in Appendix B.} formula (e.g. Feynman \& Hibbs 1965), namely
\be
{\cal K}({\bf x},a| {\bf q},0) =
\int^{{\bf x}(a)}_{\bf q} \left[{\cal D}{\bf x}(\tau) \right]
e^{-S/ 2\nu} \;,
\label{newbook}
\ee
where the {\it action} is given by
\be
S[{\bf x}] = \int_0^a d \tau {\cal L}({\bf x},
\dot {\bf x}, \tau) = \int_0^a d \tau
\left[{{\dot {\bf x}}^2\over 2} - \eta({\bf x}, \tau)
\right] \;,
\label{action}
\ee
with ${\cal L}$ the Lagrangian of a particle moving in the potential $\eta$.

In our diffusion equation the transition kernel is immediately
understood as the conditional probability of finding a particle in
${\bf x}$, at time $a$, given that it was initially in the
Lagrangian position ${\bf q}$. To better understand the
path-integral solution it is however convenient to think it in
connection with a quantum mechanical problem. It is in fact
immediate to realize that an inverse `Wick rotation' of the time
variable $a \to i a$, together with the formal replacement $2\nu
\to \hbar $ transforms eq. (\ref{fa1}) into the Schr\"odinger
equation for a particle, of unit mass, subjected to the potential
$\eta$. According to the path-integral representation of quantum
mechanics, its solution is obtained by `integrating' over all
possible paths connecting these two end-points, each one weighed
by the action $S[{\bf x}]$, calculated along this path.

Once the kernel is known, the solution of the random heat equation
with appropriate initial conditions is obtained through the
application of the Chapman-Kolmogorov equation (e.g. van Kampen 1992),
by integrating the product of the initial function ${\cal U}_0$
with the transition
kernel over the whole Lagrangian space, namely ${\cal U}({\bf
x},a) = \int d^3 q ~~{\cal U}_0({\bf q}) ~{\cal K}({\bf x},a|{\bf
q},0)$. In our case ${\cal U}_0({\bf q}) = \exp[-\Phi_0({\bf
q})/2\nu] = \exp[\varphi_0({\bf q})/2\nu]$. Thus,
\be
{\cal U}({\bf x},a) = \int d^3 q ~e^{-\Phi_0({\bf q})/2\nu}
\int^{{\bf x}(a)}_{\bf q} \left[{\cal D}{\bf x}(\tau) \right]
e^{-S/ 2\nu} \;.
\label{formalsol}
\ee

In the limit of vanishing viscosity (corresponding to the
classical limit, $\hbar \to 0$, in our quantum mechanical analog)
the dominant contribution to the path-integral comes from the
`classical' path, i.e. that which satisfies the Euler-Lagrange
equations of motion,
\be
{\partial{\cal L}\over\partial {\bf x}}-
{d\over d a} {\partial{\cal L}\over\partial \dot{\bf x}}=0
\;,
\ee
which in our case leads to Newton's second law, $\ddot {\bf x}
=-{\bf\nabla}\eta$. Thus, the particle trajectories along the
classical path read
\be
{\bf x}_{\rm cl}({\bf q}, a) = {\bf q} + a {\bf u}_0({\bf q}) -
\int_0^a \, d\tau \int_0^\tau \, d \tau^\prime {\bf \nabla}\,
\eta({\bf x}({\bf q}, \tau^\prime),\tau^\prime) \;,
\label{classtraj}
\ee
with general initial velocity ${\bf u}_0$. Note that there are
still infinitely many classical trajectories joining the two
end-points $({\bf q},0)$ and $({\bf x},a)$, corresponding to
the freedom to choose the initial velocity ${\bf u}_0({\bf q})$.

We can now expand around the classical trajectories in the
standard manner of quadratic approximations (e.g. Feynman \& Hibbs
1965), that is ${\bf x}={\bf x}_{\rm cl}+{\bf \xi}$, subject to
the constraint that the fluctuations around the classical
trajectory vanish at the end points, namely ${\bf
\xi}(0)={\bf\xi}(a)=0$. We then have
\be
S[{\bf x}]=S_{\rm cl} + \int d \tau\,\xi(\tau) \left({\delta S[{\bf
x}]\over\delta {\bf x}}\right)_{{\bf x}_{\rm cl}} + {1\over 2}
\sum_{i,j=1}^3 \int d\tau_1 \int d\tau_2 ~\xi_i(\tau_1)\xi_j(\tau_2)
\left({\delta^2 S[{\bf x}]\over \delta x_i(\tau_1)\delta x_j(\tau_2)}
\right)_{{\bf x}_{\rm cl}} \;,
\label{expand}
\ee
with the symbol $\delta$ standing for functional differentiation.
The classical action, $S_{\rm cl}$, is a function of the
end-points $({\bf q},{\bf x})$ and of the time interval $a$, which has to
be calculated by replacing the solution of the Euler-Lagrange
equations into its expression.

The action is an extremum for the classical trajectory, thus
$(\delta S /\delta {\bf x})_{{\bf x}_{\rm cl}}=0$. Inserting the
expansion (\ref{expand}) in the path-integral gives
\be
{\cal
K}({\bf x},a|{\bf q},0) = e^{-S_{\rm cl}/2\nu} \int^{\bf 0}_{\bf
0} \left[{\cal D}{\bf \xi}(\tau)\right] e^{-{1\over 4\nu}
\sum_{i,j=1}^3 \int_0^a d\tau\int_0^a d\tau_1 \int_0^a d\tau_2
~\xi_i(\tau_1)\xi_j(\tau_2)\left({\delta^2 S\over \delta x_i(\tau_1)
\delta x_j(\tau_2)}\right)_{{\bf x}={\bf x}_{\rm cl}}}  \;,
\label{twobook}
\ee
where we used the fact that the Jacobian
of the transformation from ${\bf x}$ to ${\bf \xi}$ is unity. Note
that the path-integral in (\ref{twobook}) starts and ends at zero,
because the fluctuations vanish at the end-points. Contrary to an
ordinary integral, the fact that the upper and lower limits are
the same does not imply that the path-integral vanishes. This can
be clearly seen by considering the path-integral of a free
particle (see Appendix B), which can be evaluated exactly and
does not vanish even if one is only interested in the path which
starts and ends at the same point (e.g. Feynman \& Hibbs 1965).
Since the integral over ${\bf \xi}$ starts and ends at ${\bf 0}$,
it can only be a function, $F(a)$, of the end-times. Therefore, we
can rewrite the above path-integral as ${\cal K}({\bf x},a|{\bf
q},0) = F(a) \exp[-S_{\rm cl}/2\nu]$. We need not find the
explicit expression for the pre-factor $F(a)$, since this function
cancels for the velocity. The solution of the random heat equation,
for small $\nu$, reads
\be
{\cal U}({\bf x},a) = F(a) \int d^3 q ~e^{ - \Phi_{\rm cl}({\bf
x},{\bf q},a)/2\nu} \;,
\label{step1expotential}
\ee
with
\be
\Phi_{\rm cl}({\bf x},{\bf q},a) \equiv S_{\rm cl}({\bf x},{\bf q},a) + \Phi_0({\bf q})\;.
\label{classvelpot}
\ee
Replacement into eq. (\ref{reverse}) yields
\be
{\bf u}({\bf
x},a) = {\int d^3 q ~\nabla_{\bf x} S_{\rm cl}({\bf x},{\bf q},a)
~e^{ - \Phi_{\rm cl}({\bf x},{\bf q},a)/2\nu}
\over \int d^3 q ~e^{ - \Phi_{\rm cl}({\bf x},{\bf q},a)/2\nu}} \;.
\label{step1solution}
\ee

In the case of vanishing force, eq. (\ref{step1solution}) reduces
to the well-known exact solution of the three-dimensional Burgers
equation (see Appendix B), which underlies the adhesion approximation
for the dark matter component. In our case, it represents
a useful approximation, which could be used for a numerical
evaluation of the baryon velocity field, by extending the
technique applied by Nusser \& Dekel (1990) and Weinberg \& Gunn (1990a,b). 
The main practical difficulty in using the solution (\ref{step1solution})
is, however, represented by the lack of an explicit expression for
the force $\nabla \eta$, as a function of ${\bf x}$ and $a$. A
possible shortcut would be to exploit the Lagrangian approximation
for the trajectories, introduced in Section 4.1. This method will
be discussed in Section 6.

Once the peculiar velocity field of eq. (\ref{step1solution}) is known
at each Eulerian point ${\bf x}$ as a function of time, one can
obtain the actual baryon trajectories, by numerically integrating
the integral equation (e.g. Nusser \& Dekel 1990; Weinberg \& Gunn 1990a,b)
\be
{\bf x}({\bf q},a) = {\bf q} + \int_0^a d\tau
~{\bf u} ({\bf x}({\bf q},\tau),\tau) \;,
\label{physicaltraj}
\ee
starting from Lagrangian particle positions on a grid. From these 
trajectories the baryon density can be found, according to the
standard formula $1+ \delta({\bf x},a) = 
|\!|\partial {\bf x}({\bf q},a) / \partial {\bf q}|\!|^{-1}$.

\subsection{Steepest-descent approximation}

In the limit of small $\nu$ we can evaluate the integral over
initial positions in eq. (\ref{step1expotential}) using the
steepest-descent, or saddle-point, approximation. We have
\ba
{\cal U}({\bf x},a) & = & F(a) ~e^{ -\Phi_{\rm cl}({\bf x},{\bf
q}_s,a)/2\nu} \int d^3 \delta q ~e^{-{1 \over
4 \nu} \sum_{i,j=1}^3 \delta q_i \delta q_j \left({\partial^2
\Phi_{\rm cl}({\bf x},{\bf q},a) \over\partial q_i\partial q_j}
\right)_{{\bf q}={\bf q}_s}} \nonumber \\
& = & F(a) ~(4\pi\nu)^{3/2} \sum_s ~j_s({\bf x},{\bf q}_s,a)
e^{-\Phi_{\rm cl}({\bf x},{\bf q}_s,a)/2\nu} \;,
\label{threefourbook}
\ea
where
\be
j_s({\bf x},{\bf q_s},a) =
\left({\rm det}\left[{\partial^2 \Phi_{\rm cl}({\bf x},
{\bf q},a) \over \partial q_i \partial q_j}
\right]_{{\bf q}={\bf q}_s}
\right)^{-1/2} \;.
\ee

The sum in eq. (\ref{threefourbook}) extends over the {\it
saddle points} ${\bf q}_s$, i.e. the Lagrangian coordinates
corresponding to the absolute minima of the function
$\Phi_{\rm cl}({\bf x},{\bf q},a)$, for fixed $\bf x$ and $a$.
These are found by solving the equation
\be
\nabla_{\bf q} \Phi_{\rm cl}({\bf x},{\bf
q},a)\bigg|_{{\bf q}={\bf q}_s} =
\nabla_{\bf q}\left(S_{\rm cl}({\bf x},{\bf q},a) +
\Phi_0({\bf q})\right)\bigg|_{{\bf q}={\bf q}_s}  = 0 \;.
\label{saddle}
\ee
Since, for the classical action, ${\bf \nabla}_{\bf q} S_{\rm cl}
= - {\bf u}_0({\bf q})$, we see that the classical orbit
(\ref{classtraj}) passes through the saddle point ${\bf q}_s$ if
the initial velocity satisfies ${\bf u}_0({\bf q}) = \nabla_{\bf
q} \Phi_0({\bf q})$ and $\Phi({\bf x},{\bf q},a) \geq \Phi({\bf
x},{\bf q}_s,a)$, for any ${\bf q} \neq {\bf q}_s$.

We then obtain
\be
{\bf x}({\bf q}_s, a) = {\bf q}_s + a \nabla_{\bf q} \Phi_0({\bf q}_s) -
\int_0^a d\tau \int_0^\tau
d\tau^\prime
~\nabla_{\bf x}\eta({\bf x}_{\rm cl}({\bf q}_s,\tau^\prime),
\tau^\prime) \;,
\label{trajec}
\ee
for the particle trajectory, which coincides with eq. (\ref{igmtraj}),
as $\Phi_0=-\varphi_0$.

Replacing eq. (\ref{threefourbook}) in eq. (\ref{reverse}) we
finally obtain our saddle-point solution of the forced Burgers
equation for the baryon velocity field
\be
{\bf u}({\bf x}, a) = \sum_s w_s({\bf x},{\bf q}_s,a)
\nabla_{\bf x} S_{\rm cl}({\bf x},{\bf q}_s,a)  \;,
\label{spsolution}
\ee
where the (normalized) weights read
\be
w_s({\bf x},{\bf q}_s,a) = {j_s({\bf x},{\bf q}_s,a)
e^{- \Phi_{\rm cl}({\bf x},{\bf q}_s,a)/2\nu}
\over \sum_s j_s({\bf x},{\bf q}_s,a)
e^{- \Phi_{\rm cl}({\bf x},{\bf q}_s,a)/2\nu}}
= {j_s({\bf x},{\bf q}_s,a) \over \sum_s j_s({\bf x},{\bf q}_s,a)}
\label{weight}
\ee
(the last step is justified by the fact that, for a given ${\bf x}$ and
$a$, all the absolute minima have the same $\Phi_{\rm cl}$).

Since the Eulerian gradient of the action is the velocity along
the classical trajectory, we have from eq. (\ref{classtraj})
\be
\nabla_{\bf x} S_{\rm cl}({\bf x},{\bf q}_s,a)) =
{\bf u}_{\rm cl}({\bf x}({\bf q},a), a) =
{{\bf x} - {\bf q}_s \over a}-
\int_0^a d\tau ~\nabla_{\bf x}\eta({\bf x}({\bf q}_s,
\tau),\tau) + \int_0^a d\tau \int_0^\tau
d\tau^\prime ~\nabla_{\bf x}\eta({\bf x}({\bf q}_s,
\tau^\prime),\tau^\prime) \;.
\label{classvel}
\ee

At early times, there is a single contribution to the
steepest-descent velocity field, coming from a unique Lagrangian
saddle point ${\bf q}_s$ that minimizes $\Phi_{\rm cl}$ for
given ${\bf x}$ and $a$, so that ${\bf u} ({\bf x},a) = {\bf
u}({\bf x}_{\rm cl}({\bf q}_s,a),a)$. With time passing the
mapping ${\bf q}_s \rightarrow \bf x$ becomes many-to-one. At this
stage, according to the steepest-descent solution, the flow at
$\bf x$ becomes a weighted average of the velocity of all the mass
elements converging to that point at $a$ along their classical
orbits. This sets the onset of the epoch of pancake formation in
our model. This interpretation of our solution will become more
clear from the geometrical construction outlined in the next
Section. \\

In concluding this Section, let us come back to our analogy with
the KPZ model. A different version of it, which is also often used
in condensed matter physics, is obtained by making the Hopf-Cole
nonlinear transformation on the KPZ equation, which leads to the
linear diffusion equation for the free energy ${\cal
Z}=\exp(-\Phi/2\nu)$ (Kardar \& Zhang 1987). Its path-integral
representation is then interpreted as follows. The path traced by
a particle starting from an initial point ${\bf x}(0) = {\bf q}$
and arriving at a final point ${\bf x}(a)={\bf x}$, moving in $D$
spatial dimensions, can be viewed as a {\it polymer} in $D+1$
dimensions joining $({\bf q},0)$ to $({\bf x},a)$, with the time
variable interpreted as a spatial one along the direction of main
extension of the polymer. One specifically uses the term {\it
directed} polymer to emphasize that the path is constrained to go
only forward in time without crossing itself. The case studied
here of vanishing viscosity is frequently referred to, in the
language of KPZ and directed polymers, as the `strong coupling',
or `zero-temperature', limit. This coincides with the limit of
large Reynolds number, in the language of turbulence, or to the
classical limit, in our quantum mechanical analog.

\section{Drawing the skeleton of the IGM distribution}

There are many possible ways to look for a numerical solution of
the forced Burgers equation. Interesting techniques might be
devised, based on the analogy with the KPZ model. In the $\nu \to
0$ limit, in fact, the problem of finding the `free energy' ${\cal
Z}$, corresponding to our potential $\cal U$, reduces to a simple
{\it optimization} problem. In the KPZ context, one frequently
uses a numerical scheme referred to as the {\it transfer matrix}
method (Kardar \& Zhang 1987). First, one generates a random
energy landscape (e.g. in 2D, on a square lattice, one puts random
numbers with a certain distribution and correlation, depending on
the statistical properties of the noise $\eta$). The {\it optimal}
particle starts at a lattice point $\bf q$, chooses the next
lattice point which would cost it the least energy, jumps to it
and so on, continuing its journey to the final point $\bf x$. In
this way an {\it optimal polymer}, joining $({\bf q},0)$ to $({\bf
x},a)$, is formed with the minimum amount of energy. The free
energy ${\cal Z}$ is then evaluated by weighing the optimal
polymer, which ends there, over all the possible starting points.
A 1D illustration of this numerical method is given in 
Figure 6. 

%%%%%%%%%%%%%%%%%%%%% FIGURE 8-->6 %%%%%%%%%%%%%%%%%%%%%%%%%%%%%
%%%%%%%%%%%%%%%%%%%%% FIGURE 8-->6 %%%%%%%%%%%%%%%%%%%%%%%%%%%%%
\begin{figure}
\hskip 0.8cm\epsfxsize= 14truecm{\epsfbox{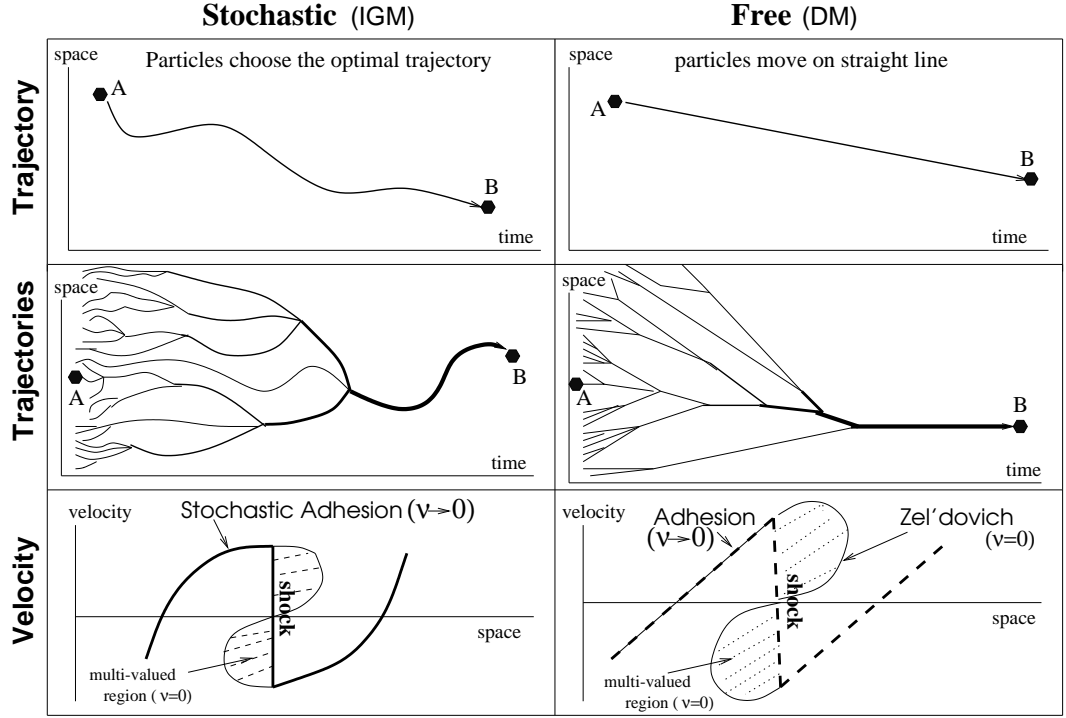}}
\caption
{
In the top figures the path of a single particle before falling into
a shock is shown for the stochastic (IGM) and free (DM) adhesion models.
The central right panel shows a collection of inertial paths
which form when particles collide, merge and move together, conserving
momentum. The left central figure shows an ensemble of paths (polymers) of
lowest energy in a random energy landscape. Each polymer has
its end points fixed and finds the optimal path in between.
The shocks of the Burgers equation are the optimal polymers in
the language of the interfacial growth.
The tree structure is a characteristic of many optimization problems.
The bottom panels show the velocity field at a fixed
large time in the three cases of zero viscosity 
(Zel'dovich), free adhesion and stochastic
adhesion. In the Zel'dovich approximation, multivalued regions
bounded by caustics
are formed since the particles can move through each other. The caustics
are replaced by shocks of finite mass once
a viscosity term stops the shell-crossing of the
particles. This holds for both free and stochastic adhesion models.
Finally, a new feature is introduced into the picture:
in the presence of a stochstic force, particles trajectories are no
longer inertial, but are curved, as shown in the bottom left panel.
}
\end{figure}
%%%%%%%%%%%%%%%%%%%%%% FIGURE 8-->6 %%%%%%%%%%%%%%%%%%%%%%%%%%%%%
%%%%%%%%%%%%%%%%%%%%%% FIGURE 8-->6 %%%%%%%%%%%%%%%%%%%%%%%%%%%%%

The application of a similar scheme to the cosmological structure
formation  problem is however complicated by the colored and
non-stationary time-correlation properties of our random
potential, as well as by the fact that the cosmological
interesting regime does not usually occur at asymptotically large
times, unlike what happens in the case of directed
polymers. \\

Let us next describe an alternative method, based on our
steepest-descent solution of the forced Burgers equation. Two main
technical difficulties arise, when one tries to implement the
saddle-point solution to simulate the baryon distribution. The
first concerns the velocity along the individual classical
trajectories, which contribute to the sum in eq.
(\ref{spsolution}): this involves a step-by-step integration of
the force along the trajectory. The second concerns the weight
$w_s$ which requires explicit knowledge of $S_{\rm cl}$ along
these classical trajectories. The problem is equivalent to that of
finding the velocity potential, $\Phi_{\rm cl}({\bf x},{\bf q},a)
= S_{\rm cl}({\bf x},{\bf q},a) + \Phi_0({\bf q})$, which solves
the `classical' Hamilton-Jacobi equation (\ref{bernoulli1}) with
initial conditions $\Phi_{\rm cl}({\bf q},{\bf q},0)=\Phi_0({\bf
q})$. The technique is therefore completely analogous to that used
in the free adhesion approximation for DM particles, but is made
more involved by the presence of the random potential $\eta$.

A substantial simplification is obtained, if we adopt the same
approximation of Section 4.1, which basically consists in
expanding the particle orbits to first order in the displacement
vector. Let us describe the method in detail. For each orbit the
system contains a cyclic variable corresponding to the initial
velocity, namely
\be
{\bf u}_0 = {{\bf x} - {\bf q} \over a} +
\nabla_{\bf q} (\psi - \varphi_0)  \;,
\label{cyclic}
\ee
where $\psi$ is the baryon potential defined in eq.
(\ref{barpot}). Replacing ${\dot {\bf x}} = {\bf u}_0 + a
\nabla_{\bf q}(\varphi_0 - \psi)$ in the action (\ref{action}) and
then using (\ref{cyclic}), we obtain, to second order in the
displacement vector ${\bf S}$,
\be
\Phi_{\rm cl}({\bf x},{\bf q},a) = {({\bf x} - {\bf q})^2
\over 2 a} - a ({\bf x} - {\bf q}) \cdot \nabla_{\bf q} {\dot
\psi}({\bf q},a) -
\psi ({\bf q},a) - a {\dot \psi}({\bf q},a)
-  {1 \over 2} \int_0^a d\tau \tau^2
\left( \nabla_{\bf q} {\dot \psi}({\bf q},\tau) \right)^2 +
{\cal O}({\bf S}^3) \;,
\label{lagrclassaction}
\ee
where, for consistency, we expanded the potential $\eta$ to first
order in the displacement vector around the Lagrangian coordinate
(as $\eta$ itself is a first-order quantity), i.e.
\be
\eta({\bf x},a) = \eta({\bf q},a) + \nabla_{\bf q} \eta({\bf
q},a) \cdot \left[ a {\bf u}_0({\bf q}) - \nabla_{\bf q}
\left(\psi({\bf q},a) - \varphi_0({\bf q})\right) \right] + {\cal
O}({\bf S}^3) \;,
\label{expandeta}
\ee
It is immediate to check that the velocity potential of eq.
(\ref{lagrclassaction}) gives the required solution of the
Hamilton-Jacobi equation (\ref{bernoulli1}), if the potential
$\eta$ in its RHS is consistently expanded as in eq.
(\ref{expandeta}). From eq. (\ref{lagrclassaction}) we can
immediately obtain an explicit expression for $j_s$ which appears
in our saddle-point solution through the weight of eq.
(\ref{weight}). We have
\be
j_s({\bf q_s},a) = \left({\rm
det}\left[{\delta_{ij} \over a} - {\partial^2 \psi({\bf q},a)
\over \partial q_i \partial q_j} + a {\partial^2 {\dot \psi}({\bf q},a)
\over \partial q_i \partial q_j} \right]_{{\bf q}={\bf
q}_s} \right)^{-1/2} + {\cal O}({\bf S}^2) \;.
\label{js}
\ee

The Lagrangian approximation for the classical trajectory
\be
{\bf x}_{\rm cl}({\bf q}_s, a) = {\bf q}_s - a \nabla_{\bf q}
\psi({\bf q}_s,a)
\label{repclasstraj}
\ee
and the corresponding velocity
\be
{\bf u}_{\rm cl}({\bf x},{\bf q}_s,a) = {{\bf x} - {\bf q}_s
\over a} - a \nabla_{\bf q} {\dot \psi} ({\bf q}_s,a)
\label{repclassvel}
\ee
are the remaining ingredients of the saddle-point solution
(\ref{spsolution}). Note that everything is now explicitly given
in terms of the linear baryon potential $\psi$ and its time
derivatives.

We are now ready to introduce a geometrical construction of the
saddle-point solution, obtained in analogy with the adhesion
approximation, which aims at obtaining the skeleton of the IGM
distribution. The graphic technique has been largely applied in
connection with the solution of the one-dimensional Burgers
equation (e.g. Burgers 1974). In the cosmological context this
method has been described in a number of papers (e.g. Gurbatov et
al. 1985, 1989; Kofman \& Shandarin 1988; Shandarin \& Zel'dovich
1989; Vergassola et al. 1994; Sahni \& Coles 1995) and applied 
in one, two and three-dimensional numerical simulations of the large-scale
structure of the Universe (Kofman, Pogosyan \& Shandarin 1990;
Williams et al. 1991; Kofman et al 1992; Sahni, Sathyaprakash \& 
Shandarin 1994; Sathyaprakash et al. 1995), finding good agreement 
with N-body simulations in models which have sufficient large-scale power
(e.g. Sathyaprakash et al. 1995).

Let us start by noting that, for given ${\bf x}$ and $a$, the classical
velocity potential $\Phi_{\rm cl}({\bf x},{\bf q},a)$
describes a random three-dimensional hypersurface in Lagrangian space.
The idea is that absolute minima of this hypersurface correspond to
Lagrangian points of first-touching with the hyperplane
$\Sigma_h({\bf q}) = h$, as the parameter $h$ is raised
from $-\infty$ \footnote{In the free adhesion case it is
actually more convenient to look for the first-touching points of the
paraboloid ${\cal P}_h({\bf q}|{\bf x},a)=({\bf q} -{\bf x})^2/2a + h$,
with the random hypersurface $\Phi_0({\bf q})$, as the latter is both
${\bf x}$ and $a$ independent.}. 
%
%%%%%%%%%%%%%%%%%%%%% FIGURE 9 %%%%%%%%%%%%%%%%%%%%%%%%%%%%%
%%%%%%%%%%%%%%%%%%%%% FIGURE 9 %%%%%%%%%%%%%%%%%%%%%%%%%%%%%
%\begin{figure}
%\begin{center}
%\epsfxsize= 14truecm{\epsfbox{paraboloid.eps}}
%\end{center}
%\caption
%{
%The paraboloid construction which has been frequently used
%to find the solutions to the adhesion model for DM is shown on the 
%left. The picture on the right demonstates how a modified scheme, based 
%on a hyperplane, rather than a paraboloid, can be used to find the position 
%of the IGM particles, which are subjected to a stochastic force.
%}
%\end{figure}
%%%%%%%%%%%%%%%%%%%%%% FIGURE 9 %%%%%%%%%%%%%%%%%%%%%%%%%%%%%
%%%%%%%%%%%%%%%%%%%%%% FIGURE 9 %%%%%%%%%%%%%%%%%%%%%%%%%%%%%

The geometrical construction can proceed along similar lines
as for the adhesion model, namely:

\noindent
{\it i)} At each time $a$, and for each point $\bf x$ in Eulerian
space, find the hyperplane $\Sigma_h({\bf q})$ tangential to
$\Phi_{\rm cl}({\bf x},{\bf q},a)$ at the Lagrangian point(s)
${\bf q}_{s_i}$, by increasing the parameter
$h$ from $-\infty$, so that the two hypersurfaces do not
cross anywhere.

\noindent
{\it ii)} Once the saddle point(s) ${\bf q}_{s_i}({\bf x},a)$ are
found, the corresponding velocity at $\bf x$ at time $a$ is given by
eq. (\ref{spsolution}), as the weighted sum of single-particle velocities
(\ref{repclassvel}), with the weights given by eq. (\ref{weight}), with $j_s$
from eq. (\ref{js}) and $S_{\rm cl}$ from eq. (\ref{lagrclassaction}).

At early times, the hypersurface $\Phi_{\rm cl}$ is a narrow, distorted
paraboloid whose apex starts in ${\bf q}={\bf x}$ and gradually moves away
from it. At these early times, the curvature of this random hypersurface
is large, so that there is a one-to-one correspondence between Lagrangian and
Eulerian space. As time passes, our classical velocity potential becomes
shallower, so that it becomes increasingly frequent to find points
$\bf x$ such that the first-touching condition of $\Sigma_h$ and
$\Phi_{\rm cl}$ occurs at multiple saddle points ${\bf q}_{s_i}$. The
Lagrangian to Eulerian correspondence has now become many-to-one,
with the degree of the mapping being inversely proportional to the
dimensionality of the structure formed. Two simultaneous points of
touching, ${\bf q}_{s_1}({\bf x},a)$ and ${\bf q}_{s_2}({\bf
x},a)$, signal the formation of a sheet, or pancake, in the
corresponding Eulerian location $\bf x$, three indicate the
presence of a filament, and four that of a knot. As is well known
from the adhesion model, filaments occur at the intersection of
pancakes, while nodes at the intersection of filaments.

There is only a caveat in our geometrical construction, owing to
the approximate form of eq. (\ref{lagrclassaction}). Because of
this, eq. (\ref{saddle}) is only approximately satisfied, namely
$\nabla_{\bf q} \Phi_{\rm cl}$ calculated from eq.
(\ref{lagrclassaction}) contains spurious terms which are of
second-order in the displacement vector. As a consequence, our
accuracy in the determination of the saddle points is limited to
first order. Nevertheless, our Lagrangian approximation is likely
to be quite accurate not only at early times, when the evolution
is linear, but also at later times, as far as the skeleton of the
IGM structure is concerned. In fact, the dominant contribution to
our saddle-point solution in denser regions, such as pancakes,
filaments and knots, is expected to come from the deepest
potential wells [see, e.g., the related interpretation of the
path-integral solution by Zel'dovich et al. (1985, 1987)], i.e.
precisely from those Lagrangian sites where the particles
experience the smallest displacement from their initial positions
(e.g. Matarrese et al. 1992). \\

As for the standard adhesion model, the above geometrical
construction cannot describe the internal structure of dense
regions, like pancakes, filaments and knots, as it is only
designed to find the sites where these structure form and to
follow the subsequent evolution through a continuous merging
process (e.g. Shandarin \& Zel'dovich 1989). From this point of
view, it is clear that a numerical algorithm based on the direct
integration of the solution in eq. (\ref{step1solution}), would be
far more suitable to follow the baryon distribution on smaller
scales. This kind of numerical integration can be performed by
extending the technique used by Weinberg \& Gunn (1990a,b) and
Melott, Shandarin \& Weinberg (1994). For the free adhesion
approximation, Weinberg \& Gunn notice that the exact solution
[our eq. (\ref{threethreebook})] appears as a Gaussian convolution
integral, which is easy to handle numerically. For our stochastic
adhesion model, this type of calculation is obviously made more
complex by the presence of the random force term, which does not
generally allow to get an exact explicit expression for the
action. This problem, however, is enormously simplified if one
adopt the approximate expression of eq. (\ref{lagrclassaction}),
for the classical action. As mentioned previously, the last step
in this scheme would consist in the numerical integration of eq.
(\ref{physicaltraj}) to obtain the actual particle positions.
Although the computational time required to perform the
integration in eq. (\ref{step1solution}) plus that of eq.
(\ref{physicaltraj}) implies only a modest speed up, compared to a
particle-mesh N-body code; one should keep in mind that our stochastic
adhesion model actually aims at describing a physical situation
where the exact treatment of the two-fluid evolution would require a
full hydrodynamical code.

\section{Statistics of the density field}

The stochastic adhesion model provides a direct insight on the
properties of the velocity field away from the linear regime. A
vast literature is devoted to the study of the statistical
properties of this field, in cases when the external potential
$\eta$ is a Gaussian random field which is either time-independent
(e.g. Zel'dovich et al. 1987), or with white-noise time-correlation 
properties (e.g. Bouchaud et al. 1995; E et al. 1997). It
is generally found that the system develops intermittency at late
times. Based on this property, Jones (1999) argues that the
velocity field is normally  distributed even during the nonlinear
stage. He also suggests that this may provide a dynamical motivation
for a Lognormal PDF of the density field in the baryonic component. 

More complex is to analyze the statistical behavior of the
velocity and density field in our model, owing to the non-trivial
time-correlation properties of the noise, as well as to the fact
that we are interested in the behavior of the system in the
intermediate regime, when most of the matter still resides in pancakes
and filaments, rather than having relaxed in the most massive
clumps, corresponding to the nodes of the cellular structure. We
will only give here a preliminary introduction to this problem,
focusing on the density field, rather than the peculiar velocity.

A convenient form of the comoving local matter density can be
obtained by integrating the continuity equation along the particle
trajectories. One gets (e.g. Matarrese et al. 1992)
\be
1 + \delta({\bf x}({\bf q},a),a) = \exp\left[-
\int_0^a d \tau ~\theta({\bf x}({\bf q},\tau),\tau) \right] \;,
\label{integrdensity}
\ee
where $\theta({\bf x},a) \equiv \nabla_{\bf x} \cdot {\bf u}({\bf
x},a)$ and the time integration is along the particle trajectory.
The validity of this integral expression for the density is restricted
to the single-stream regime. We are then allowed to apply it in our case,
provided the viscosity coefficient $\nu$ takes a finite value. For each
Eulerian position ${\bf x}$ at time $a$ there is, in fact, a unique
Lagrangian element ${\bf q}$ implicitly defined through eq.
(\ref{physicaltraj}).

A relevant statistical information on the density field could
then be obtained by studying its ({\it disconnected}) {\it moments},
$\langle (1 + \delta)^p\rangle$. We have
\be
\bigg\langle \bigl(1 +
\delta({\bf x},a)\bigr)^p\bigg\rangle_{\rm E} = 
\bigg\langle \exp\left[- (p-1)
\int_0^a d \tau ~\theta\bigl({\bf x}({\bf q},\tau),\tau\bigr) 
\right] \bigg\rangle_{\rm L} \;,
\label{moments}
\ee
where the subscripts `E' and `L' denote the statistical ensemble
over which the average is performed: the one on the LHS involves
the Eulerian PDF, whereas that on the RHS is most conveniently
obtained with the Lagrangian PDF. To account for this difference
we multiplied the RHS by the Jacobian $|\!|\partial {\bf x}
/\partial {\bf q}|\!|=(1+ \delta)^{-1}$ (e.g. Kofman et al. 1994).
Note that the Lagrangian-to-Eulerian mapping is always one-to-one,
as the velocity field solving the forced Burgers equation is
non-singular everywhere for any finite value of $\nu$; thus the
actual fluid elements, which move according to eq.
(\ref{physicaltraj}), do not experience orbit-crossing during
their evolution.

The main technical difficulty in dealing with eq. (\ref{integrdensity}),
or any equivalent expression for the density field, resides in the
inversion of the {\it Lagrangian map} ${\bf q} \to {\bf x}({\bf q},a)$
(e.g. Vergassola et al. 1994). The same problem, of course, makes it
generally difficult to deal analytically with its statistical properties
(such as the disconnected moments above).
An alternative technique to reconstruct the matter density field from
the velocity, solving the Burgers equation, is discussed by Gurbatov
(1996; see also Gurbatov, Malakhov \& Saichev 1991), who propose a
modification of the continuity equation to simplify this inversion
problem. Another interesting approach to the same problem is
adopted by Vergassola et al. (1994), who implement a Fast Legendre
Transform algorithm.
Here we will sketch the first steps of a different method, which aims at 
relating the mass density in ${\bf x}$ at time $a$ to the properties of 
the classical velocity field at the saddle-points ${\bf q}_s ({\bf x},a)$.

\subsection{The velocity divergence}

The first step to obtain the density field is to write a suitable
expression for the velocity divergence $\theta$.
Using eq. (\ref{reverse}), we can write the velocity divergence in
the general form
\be
\theta = {{\bf u}^2\over 2 \nu} -
2 \nu {\nabla^2 {\cal U} \over {\cal U}} \;,
\label{veldiv}
\ee

Next, we can replace for ${\bf u}$ and ${\cal U}$ the
expressions of eq. (\ref{step1solution}) and (\ref{step1expotential})
(which would be exact in the free adhesion case), which gives
\be
\theta({\bf x},a) = {3 \over a} - {1 \over 2 \nu}
\bigl[\!\bigl[ \bigl({\bf u}_{\rm cl}({\bf x},{\bf q},a) -
\bigl[\!\bigl[ {\bf u}_{\rm cl}({\bf x},{\bf q},a) \bigr]\!\bigr]
\bigr)^2 \bigr]\!\bigr] \;,
\label{theta}
\ee
where we introduced the symbol
\be
\bigl[\!\bigl[ {\bf A}({\bf x},{\bf q},a)\bigr]\!\bigr] 
\equiv {\int d^3q ~{\bf A}({\bf x},{\bf q},a)
~e^{- \Phi_{\rm cl}({\bf x},{\bf q},a)/2\nu} \over
\int d^3q ~e^{- \Phi_{\rm cl}({\bf x},{\bf q},a)/2\nu}} \;,
\ee
for any scalar, vector or tensor ${\bf A}$.
To get the first term on the RHS of eq. (\ref{theta}) we used the
approximate form for the classical action deriving from eq.
(\ref{lagrclassaction}).

To better understand the meaning of the expression (\ref{theta})
for the velocity divergence, we can consider the small $\nu$ limit,
using the steepest descent approximation. We get
\be
\theta({\bf x},a) = \sum_s w_s({\bf q}_s,\tau) ~\theta_{\rm cl}
({\bf x}({\bf q}_s,a),a) - {1 \over 2 \nu}
\delta {\bf u}^2({\bf x},a) \;,
\label{steeptheta}
\ee
where $\theta_{\rm cl} \equiv \nabla_{\bf x}
{\bf u}_{\rm cl} = \nabla^2_{\bf x} \Phi_{\rm cl}$, and
\be
\delta {\bf
u}^2({\bf x},a) = \sum_s w_s({\bf q}_s,a)
{\bf u}_{\rm cl}^2({\bf q}_s,a) - \sum_s \sum_{s^\prime}
w_s({\bf q}_s,a) w_{s^\prime}({\bf q}_{s^\prime},a)
{\bf u}_{\rm cl}({\bf q}_s,a)
{\bf u}_{\rm cl} ({\bf q}_{s^\prime},a) \;,
\label{veldisp}
\ee
from which it becomes evident its role as {\it mean square velocity
dispersion} caused by the convergence of several classical streams in
${\bf x}$. Note that, by definition, $\delta {\bf u}^2\geq 0$.
In order to obtain the expression of eq. (\ref{steeptheta}), we
expanded the classical velocity to first order around the saddle point,
\be
u^i_{\rm cl}({\bf x},{\bf q},a) = u^i_{\rm cl}({\bf x},{\bf q}_s,a)
- {1 \over a} \left(\delta^i_j + a^2 {\partial^2 {\dot \psi}({\bf q}_s,a)
\over \partial q_i \partial q^j} \right)\left(q^j - q^j_s\right)
\ee
and wrote the classical velocity divergence in the form
\be
\theta_{\rm cl}({\bf x}({\bf q},a),a) = {3 \over a} - {1 \over a}
\left[\delta_{ij} + a^2 {\partial^2 {\dot \psi}({\bf q},a)
\over \partial q^i \partial q^j}\right]
\left[ \delta^{ij} - a {\partial^2 \psi({\bf q},a)
\over \partial q_i \partial q_j} \right]^{-1} \;,
\ee
which follows from taking the Eulerian divergence of eq. (\ref{repclassvel}).
Note that, while in the free adhesion case, eq. ({\ref{steeptheta}) is
the direct result of the saddle-point approximation, in our stochastic
adhesion model, that expression has been derived only to
first order in the displacement vector. Nonetheless, we believe that
the validity of eq. ({\ref{steeptheta}) is much more general than its
approximate derivation might suggest.

\subsection{Evolution of the density field}

We are now in a position to discuss the evolution of the density field
in the various stages of the structure formation process.

\subsubsection{Linear stage}

At early times, when the system is well approximated by Eulerian
first-order perturbation theory, $\int_0^a d \tau \theta \approx -
a \nabla^2 \psi \ll 1 $, and the density fluctuation is
linearly related to that of the initial gravitational potential,
which we assumed to be a uniform Gaussian process. Density
fluctuations at this time are also Gaussian distributed.

\subsubsection{Weakly nonlinear stage}

Later on, the system enters a weakly nonlinear regime, when
density fluctuations are no longer small, but
caustic formation is still a sporadic event. In this laminar stage
the evolution of the system is generally well described by
Lagrangian first-order perturbation theory (see Section 4.1). Our
solution (\ref{steeptheta}) yields
\be
1 + \delta({\bf x}({\bf q},a),a) = \exp \left( -
\int_0^a d\tau ~\theta_{\rm cl}({\bf x}({\bf q}_s, \tau),\tau)
\right) \;,
\label{lamdens}
\ee
as, for each $\bf x$, there is a unique steepest-descent solution,
corresponding to the particle, with initial coordinate ${\bf q}_s$,
which reaches that point at time $a$. This expression is of course
equivalent to the more conventional one of Section 4.1.2 [e.g.
Appendix A in (Matarrese et al. 1992)]. At this early nonlinear
stage, the density distribution is no longer Gaussian. Obtaining
the PDF of the mass density at this stage is a classical problem
which has been solved long ago by Doroshkevich (1970), and
analyzed in more detail by several authors (e.g. Kofman et al. 1994;
Bernardeau \& Kofman 1994), for the collisionless case. As
anticipated in Section 4.1.2, the presence of non-zero
acceleration in the motion of our collisional fluid elements does
not imply relevant modifications compared with the DM
case, so, for instance, the functional form of the PDF is 
unchanged.

\subsubsection{Mildly nonlinear stage}

Let us finally come to the most interesting issue: the
analysis of the mildly nonlinear stage, when dense, spatially
extended structures have grown in the system. It is this process
which is accompanied by the occurrence of strong phase coherence
both in the velocity and density fields.
According to our stochastic adhesion model, the formation of
a shock singularity in ${\bf x}$ at time $a$ is accompanied by the
occurrence of multiple saddle-point solutions ${\bf q}_s({\bf x},a)$.
As a consequence of these multiple solutions, the velocity dispersion
term of eq. (\ref{veldisp}) will suddenly become non-zero.
Because of the `diverging' (as $\nu \to 0$) factor $1/2\nu$, however,
any non-zero $\delta {\bf u}^2$ would signal a `singularity'
(for $\nu \to 0$) in the local density. One has, therefore, the
following picture: as $\nu \to 0$ the mass tends to get concentrated
in the cellular structure, which becomes a thin cobweb of delta-like
sheets, filaments and nodes. On the other hand, as soon as the
Lagrangian-to-Eulerian mapping becomes many-to-one our
expression (\ref{integrdensity}) for the mass density will lose
its validity, and a more clever technique should be used to account
for the actual mass converging into these nonlinear structures.
A possible hint in this direction might be given by the calculation
of the velocity field on a pancake, made by Kofman (1989),
within the free adhesion model. A more detailed analysis of this problem will
be presented elsewhere.

\section{Conclusions}

In this paper we have proposed a new dynamical model for the
formation of mildly nonlinear structure in the IGM, based on a
`stochastic adhesion' approximation to the Navier-Stokes
equation, which governs the evolution of the collisional baryon
component. The main idea of our model is that a random external
force can be used to approximate the composition of Hubble drag,
local gravitational force and pressure gradients. Such a force,
consistently calculated from first-order perturbation theory, has
two main effects. First, it produces a slight distortion of the
large-scale filamentary cosmic web that characterizes
the IGM distribution, compared with the DM one. Second, it adds a
small-scale noise to the otherwise deterministic evolution of the
fluid, thereby imprinting characteristic ripples on top of the
cellular network. This small-scale roughness, which is the memory
of acoustic oscillations in the baryon fluid, can be used as a
probe of the IGM equation of state and thermal history 
after hydrogen reionization.

This paper was mostly devoted to the presentation of the
stochastic adhesion model. A few applications were shown, only for
the simplified inviscid ($\nu=0$) case, where the trajectories
followed by patches of the collisional fluid are obtained through
a first-order Lagrangian approximation scheme, somehow similar to
that adopted by Gnedin \& Hui (1996) and Hui, Gnedin \& Zhang
(1997). Already at this level, however, the precise IGM thermal
history is found to affect the resulting clustering properties of
the gas, through the detailed structure of denser regions. The
second part of our paper was devoted to the solution of the forced
Burgers equation, which governs the evolution of the IGM peculiar
velocity field in our model. Among our main results is the
saddle-point solution, eq. (\ref{spsolution}), and the Lagrangian
approximation of Section 6: both are completely new and might have
much wider applications than the present context. We also sketched
a geometrical representation of our saddle-point solution,
analogous to that used for the standard adhesion model, which
allows to draw the skeleton of the IGM distribution.

We did not address here the important issue of whether pancakes or
filaments provide the dominant structure in the large-scale matter
distribution. According to the cosmic web description of Bond et al.
(1996) the nonlinear evolution of the DM brings a filamentary network
into relief. Bond \& Wadsley (1997) applied similar ideas to the IGM
distribution probed by the Ly$\alpha$ forest. Quite recently,
Colombi, Pogosyan \& Souradeep (2000) adopted topological descriptors
to conclude that the structure of the Universe at the percolating
threshold is predominantly filamentary, which confirms the results
of a percolation analysis of N-body simulations 
by Sathyaprakash, Sahni \& Shandarin (1996).
In both the free and stochastic adhesion models, this aspect of
the emerging structures on large scales is
expected to be strictly related to the stage of dynamical evolution of the
system, as well as to the value of the viscosity coefficient $\nu$
and to the resolution scale at which the structure is analy

The formation of a cellular structure on large scales, appearing
as a cobweb of interconnected sheets, filaments and nodes, both
for the DM and IGM components, is accompanied by the growth of specific 
phase coherence in the velocity and density field. We have shown how
this phenomenon, which has been dubbed `intermediate
intermittency' (Zel'dovich et al. 1985, 1987), is most naturally
described in terms of the forced Burgers equation, which underlies
our model. A wavelet analysis of the
Ly$\alpha$ forest, such as that recently initiated by Meiksin
(2000) and Theuns \& Zaroubi (2000), would be the ideal tool for a
statistical analysis of this phenomenon, even on those small
scales where the effects of the thermal gas pressure are expected
to imprint characteristic features on the baryon distribution,
which should tell us about the equation of state and detailed
thermal history of the IGM.

In concluding this paper let us emphasize that the stochastic adhesion model 
presented here might represent the ideal tool to study the statistical 
properties of the low-column density Ly-alpha forest, because of its ability 
to accurately describe the IGM distribution from the largest scales, 
poorly probed by present-day hydro-simulations, down to those regions 
which have undergone mildly non-linear evolution (baryon density contrast 
up to $5 -- 10$), which cannot be properly represented by smoothed 
versions of the Zel'dovich approximation.

\section*{Acknowledgments.} We would like to thank U. Frisch,
L. Moscardini, A. Nusser, A. Stella and M. Viel
for enlightening discussions and for technical help. RM
acknowledges the TMR Network FMRXCT980183 for financial support.

\appendix

\section{First-order Lagrangian perturbation theory for the baryons}

We will briefly derive here an expression for the baryon
trajectories within first-order Lagrangian perturbation theory.
Lagrangian approximation schemes have been successfully applied to
the weakly nonlinear evolution of the DM fluid (e.g. Buchert 1989,
1992; Moutarde et al. 1991; Bouchet et al. 1992; Catelan 1995).
Adler \& Buchert (1999) have recently applied the Lagrangian
perturbation technique to the case of a collisional
self-gravitating fluid, corresponding to $f_{\rm b}=1$ in our
equation (\ref{twocomp}). The case discussed here is instead that
of a collisional fluid moving in the gravitational field caused by
a DM component, corresponding to $f_{\rm b}=0$ in our equation
(\ref{twocomp}).

The particle trajectories of both components can be represented, before
shell-crossing, in the general form
\be
{\bf x}({\bf q},a)={\bf q}+{\bf S}({\bf q},a)
\ee
Mass conservation (starting from a uniform distribution) implies
\be
1+\delta({\bf x}({\bf q},a),a)
= {1\over {\rm det}(\partial x_\alpha/\partial q_\beta)}
= {1\over J}= 1-{\bf\nabla}_q\cdot {\bf S} +O(S^2) \;,
\label{continental}
\ee
where $J$ is the Jacobian determinant of the transformation from
Lagrangian to Eulerian coordinates.

Taking the divergence of the baryon Euler equation we obtain
\be
\nabla_{\bf x} \cdot
{\ddot{\bf S}}_{\rm b}= -{3\over 2a}
\left[ \nabla_{\bf x} \cdot {\dot {\bf S}}_{\rm b}
+ {\delta_{\rm DM} \over a}
+ {1\over (\gamma-1) a k_J^2} \nabla^2_{\bf x}
\left(1 + \delta_{\rm b}\right)^{\gamma-1}
\right]
\ee
Substituting for the overdensities from eq. (\ref{continental})
we have
\be
\nabla_{\bf x} \cdot {\ddot{\bf S}}_{\rm b}+
{3\over 2a} \nabla_{\bf x} \cdot {\dot {\bf S}}_{\rm b}
= -{3\over 2 a^2} {1-J_{\rm DM}\over J_{\rm DM}}
-{3\over 2 (\gamma-1)a^2 k_J^2}
\nabla^2_{\bf x} \left({1-J_{\rm b}\over J_{\rm b}}\right)^{\gamma-1}
\ee
Next, we expand everything to first order in the displacement
vector ${\bf S}$ and obtain
\be
\nabla_{\bf q} \cdot {\ddot {\bf S}}_{\rm b}
+{3\over 2a} \nabla_{\bf q} \cdot {\dot {\bf S}}_{\rm b}
-{3\over 2a^2} \nabla_{\bf q} \cdot {\bf S}_{\rm DM}
= {3\over 2 a^2 k_J^2} \nabla^2_{\bf q} \nabla_{\bf q}
\cdot {\bf S}_{\rm b} \;.
\label{sec}
\ee
Note that at a given time, the same Eulerian position ${\bf x}$
is generally reached by the two components (DM and baryons) starting
from different Lagrangian positions ${\bf q}_{\rm DM}$ and ${\bf q}_{\rm b}$.
To lowest order in the displacement vector, however, we are allowed to
ignore this difference and set simply ${\bf q} = {\bf q}_{\rm DM} =
{\bf q}_{\rm b}$.

We now assume that the flow is irrotational $\nabla_{\bf x} \times
{\dot {\bf S}} =0$, which, to lowest order implies ${\bf S} =
\nabla_{\bf q} \Psi$, and  reduces (\ref{sec}) to the scalar
differential equation
\be {\ddot \Psi}_{\rm b} + {3\over 2a}{\dot
\Psi}_{\rm b} - {3\over 2a^2}{1\over k_J^2} \nabla^2_q \Psi_{\rm
b}= {3\over 2a^2} \Psi_{\rm DM} \;,
\ee
which can be solved in Fourier space. Using the well-known DM
solution $\Psi_{\rm DM} = - a \varphi_0$, we obtain
\be
{\ddot
\Psi}_{\rm b} + {3\over 2a}{\dot \Psi}_{\rm b} + {3\over 2a^2}{k^2
\over k_J^2} \Psi_{\rm b}= - {3\over 2a} \varphi_0 \;,
\ee
which can be solved in exactly the same manner as we previously
solved (\ref{astrid}) to obtain the baryon overdensity. The
solution can be expressed in the form $\Psi_{\rm b} = - a
\psi_{\rm b}$, where $\psi_{\rm b}({\bf k},a) = W_{\rm b}(k,a)
\varphi_0({\bf k})$ with $W_{\rm b}$ the IGM linear filter
function obtained in the main text.

\section{Path-integral solution for the free adhesion model}

In the case of zero potential $\eta$, eq. (\ref{stochadh})
reduces to the Burgers equation.
By dropping the potential $\eta$ we obtain the free particle trajectory
\be
{\bf x}({\bf q},a) = {\bf q}+ a {\bf u}_0({\bf q})
\label{presteepest}
\ee
and the classical action reduces to
\be
S_{\rm cl}({\bf x},{\bf q},a) = \int^a_{0} d \tau {{\dot{\bf x}}^2({\bf q},
\tau) \over 2} = { a {\bf u}_0^2({\bf q}) \over 2}
= {({\bf x} - {\bf q})^2\over 2 a} \;.
\ee
We therefore have
\be
{\cal U}({\bf x},a)=F(a)
\int d^3 q
~e^{-(S_{\rm cl}({\bf x},{\bf q},a)+\Phi_0({\bf q}))/2\nu}
= {1\over (4\pi \nu a)^{3/2}}
\int d^3 q ~e^{ - {({\bf x}-{\bf q})^2
\over 4 \nu a } - {\Phi_0({\bf q})/ 2\nu} } \;,
\label{threethreebook}
\ee
where the pre-factor is evaluated by performing
the Gaussian integration over $\xi$ in eq. (\ref{twobook}).

The above expression (\ref{threethreebook}) is equivalent to
the standard expression for the adhesion approximation
(e.g. Shandarin \& Zel'dovich 1989).

In the steepest descent approximation, valid in the limit of
vanishing viscosity, the particle trajectories are given by
the solution of
\be
\nabla_{\bf q}\left( \Phi_0 +  S_{cl} \right)=0 \;,
\ee
that is
\be
\nabla_{\bf q}\Phi_0({\bf q}) + {{\bf x}-{\bf q}\over a} =0 \;,
\ee
which coincides with eq. (\ref{presteepest}), for our set of initial
conditions.
The results of this Appendix also show that the quadratic approximation
is exact for free particles.

\end{document}